%% file: main.tex
\documentclass[10pt]{article}
\usepackage[utf8]{inputenc}
\usepackage[T1]{fontenc}
\usepackage{graphicx} 
\usepackage[version=3]{mhchem}
\usepackage{subcaption}
\usepackage{booktabs}
\usepackage{comment}
\usepackage{colortbl}
\usepackage[font=small,labelfont=bf]{caption}
\usepackage[a4paper,margin=2.5cm]{geometry}
\usepackage{authblk}
\usepackage[dvipsnames]{xcolor}
\usepackage[style=numeric, sorting=none, giveninits=true, maxnames=4]{biblatex}
\usepackage{hyperref}
\hypersetup{colorlinks=true,linkcolor=blue!50!black,citecolor=blue!50!black,urlcolor=blue!50!black}
\usepackage{cleveref}
\usepackage{graphicx}
\usepackage{tabularx}
\usepackage{physics}
\usepackage{enumitem}
\usepackage{amssymb}
\usepackage{multirow}
\usepackage{braket}

\usepackage{listings}
\usepackage{minted}
\usepackage{chemformula}
\usepackage{xcolor}
\usepackage{tablefootnote}
\usepackage{multicol}
\usepackage{textcomp}
\usepackage{algorithmic}
\usepackage{tabularray}
\usepackage[]{algorithm2e}
\usepackage[normalem]{ulem}
\usepackage{makecell}
\graphicspath{{Pictures/}}

\usepackage{amssymb}
\usepackage{booktabs}

\newcommand{\teal}[1]{#1}
\newfloat{code}{tbp}{loc} 
\floatname{code}{Listing}
\setminted{fontsize=\footnotesize}\usepackage[most]{tcolorbox}
\title{AEGISS - Atomic orbital and Entropy-based Guided Inference for Space Selection - Semi-automated active space selection workflow for quantum chemistry and quantum computing applications}

\author[1,2]{Fabio Tarocco}

\author[1]{Pi A. B. Haase}

\author[1]{Fabijan Pavo\v{s}evi\'{c}}

\author[3,4]{Vijay Krishna}

\author[5]{Leonardo Guidoni}

\author[1]{Stefan Knecht}

\author[1,6]{Martina Stella\thanks{\texttt{martina.stella@algorithmiq.fi}}}

\affil[1]{Algorithmiq S.r.l., Via della Chiusa 15 20123, Milano, Italy}

\affil[2]{Department of Information Engineering, Computer Science and Mathematics,
University of L'Aquila, Via Vetoio, L'Aquila 67100, Italy}

\affil[3]{Department of Biomedical Engineering,
Cleveland Clinic Research, Cleveland, OH 44195, USA}

\affil[4]{Department of Biomedical Engineering,
Case Western Reserve University, Cleveland, OH 44106, USA}

\affil[5]{Department of Physical and Chemical Sciences,
University of L'Aquila, Via Vetoio, L'Aquila 67100, Italy}

\affil[6]{Condensed Matter and Statistical Physics,
The Abdus Salam International Centre for Theoretical Physics (ICTP),
Trieste 34151, Italy}

\begin{document}

\maketitle

\begin{abstract}
In this work, we present AEGISS, an active-space selection framework inspired by both the AVAS (Atomic Valence Active Space) and AutoCAS methodologies. By combining orbital entropy analysis with atomic-orbital projections, AEGISS enables a scalable and semi-automated workflow for constructing compact, chemically meaningful active spaces suitable for strongly correlated electronic-structure problems, substantially reducing the dependence on user intuition and trial-and-error procedures. The method is assessed on a diverse set of benchmark and application-driven systems, including Benzene, Ferrocene, the thermally activated delayed fluorescence (TADF) emitter DOBNA, and a series of Ru(II) complexes of increasing size and complexity. Across all case studies, AEGISS consistently identifies active spaces that capture the essential physics while maintaining computational efficiency, highlighting its potential as a unified active-space selection strategy for both classical multi-reference calculations and quantum algorithms.

\end{abstract}

\section{Introduction}
\label{sec:intro}
Conventional electronic-structure methods are often limited by the trade-off between computational cost and accuracy, particularly for systems exhibiting strong electron correlation or complex excited-state behavior. Density Functional Theory (DFT)~\cite{Hohenberg1964,Kohn1965} provides an excellent balance between efficiency and accuracy for many molecular and condensed-phase applications~\cite{DFT_materials,DFT_delta}. However, as a single-reference approach, DFT can struggle in situations where several near-degenerate electronic configurations contribute significantly to the wavefunction, such as transition-metal complexes, open-shell species, and electronically excited states~\cite{burke2012,Lischka2018}.

Wavefunction-based methods provide a systematic route to treating electron correlation beyond the mean-field approximation. Configuration Interaction (CI) approaches improve upon a reference determinant through a linear expansion of excited configurations, culminating in Full CI (FCI), which is exact within a finite orbital basis but computationally prohibitive because of its exponential scaling. A practical alternative is to restrict the treatment of correlation to an active space, defined by the numbers of correlated active electrons and active orbitals. Complete Active Space CI (CASCI) applies the FCI formalism within this reduced space, retaining exponential scaling with active-space size while extending the range of accessible systems. Complete Active Space Self-Consistent Field (CASSCF)~\cite{Roos1980,Andersson1990} further optimizes the active orbitals and provides a standard framework for describing strong static correlation, while post-CASSCF methods such as Complete Active-Space Second-order Perturbation Theory (CASPT2)~\cite{Andersson1992} and N-Electron Valence state Second-Order Perturbation Theory  (NEVPT2)~\cite{angeli2001,angeli2001_2,angeli2002} recover additional dynamic correlation and remain among the most important tools in modern multi-reference quantum chemistry~\cite{szal12}. In parallel, Coupled Cluster (CC) methods~\cite{crawford2000introduction,bartlett2007coupled} provide highly accurate treatments of dynamic correlation through an exponential cluster-operator ansatz, but are generally limited by their single-reference character~\cite{lyakh2012multi-reference}. Consequently, systems exhibiting strong static correlation often require multi-configurational approaches based on an appropriate choice of active space.

Despite recent advances in algorithms and high-performance computing, the exponential cost (as the number of active orbitals grows) of multi-configuration methods severely limits their applicability to relatively small active spaces~\cite{vogiatzis2017pushing}, underscoring the need for novel, scalable approaches to accurately treat strongly correlated quantum systems~\cite{knec16a}. Some relevant  examples are Density Matrix Renormalization Group (DMRG)~\cite{baia20a}, Full-CI Quantum Monte-Carlo (FCIQMC)~\cite{boot09}, (semi-stochastic) Heat-Bath CI~\cite{liju18a}, Adaptive Sampling CI (ASCI)~\cite{tubm16a}, Iterative Configuration Expansion (ICE)~\cite{chilkuri2021comparison}, and Iterative Configuration Interaction (iCI)~\cite{Hoffman2016, Liu2020}, to name just a few.

With the advent of quantum computing, multi-configuration methods and quantum algorithms~\cite{cao2019quantum,bauer2020quantum}, particularly those based on variational techniques~\cite{tilly2022variational}, are poised to work synergistically, combining the best of both worlds. \teal{Quantum computers inherently posses the potential to excel at handling} the large, entangled multi-reference wavefunctions, which are potentially computationally prohibitive for classical approaches~\cite{annurev2025}. For this reason they hold significant promise for revolutionizing quantum chemistry owing to the ability to simulate molecular systems beyond the reach of traditional computational methods~\cite{Liu2022,Cao2019,weidman2024quantum,alexeev2025perspective,doi:10.1021/acsinfocus.7e9012}.

Current quantum hardware still imposes significant limitations, such as a restricted number of qubits and sensitivity to noise and decoherence, which in turn constrain the size of active spaces that can be realistically simulated. As in classical CAS approaches, tackling larger and more complex molecular systems therefore requires selecting chemically meaningful active spaces and mapping them onto qubits, for example via transformations like Jordan–Wigner~\cite{jordan1928pauli}\nocite{bk_2002,parity_2017}. While this fermion-to-qubit mapping is a key component of quantum simulations in electronic structure, we do not discuss it in detail here and instead refer the reader to recent literature~\cite{Miller2023,Miller2024}. A short conceptual overview is included in the Supplementary Information (Section A).

A major challenge, shared by both classical and quantum multi-configuration methods, is the choice of an appropriate active space. This involves balancing chemical accuracy against computational cost, and remains a difficult and often heuristic process, prone to trial-and-error and inefficiencies. To address this issue, substantial effort has gone into developing more systematic and less empirical selection strategies. These range from simple approaches based on orbital energies (e.g. around the Fermi level) to more sophisticated methods that incorporate additional metrics. For instance, Natural Orbital Occupation Numbers (NOONs) can be used to identify relevant frontier orbitals. Other approaches include Atomic Valence Active Space (AVAS)~\cite{Sayfutyarova_2017_AVAS}, which selects orbitals based on their atomic character, and AutoCAS~\cite{Stein2016,Stein2017,Stein2019}, which relies on correlation measures from approximate wavefunctions. We discuss these methods in more detail in Section \ref{ssec:as_methods}.

A robust and efficient active-space selection strategy is essential for expanding the applicability of multi-reference electronic-structure methods and enabling practical quantum simulations of chemically relevant systems. In addition, such a capability would improve the accessibility of quantum algorithms in areas including drug discovery, materials science, and environmental chemistry, where accurate descriptions of excited states and complex molecular interactions are often essential. In this work, we introduce AEGISS, a practical workflow that combines entanglement information with chemically targeted AO projections. Inspired by key concepts underlying both AVAS and AutoCAS, AEGISS integrates these complementary ideas into a protocol designed to be robust and efficient for generating compact active spaces for classical and quantum calculations.

The paper is organized as follows. Section~\ref{sec:background} and Section~\ref{ssec:as_methods} introduce the challenges of multi-reference electronic structure methods and review existing active space selection strategies. Section~\ref{ssec:workflow} presents our workflow, AEGISS, and details each step. To assess its performance and flexibility, we apply AEGISS to five case studies. The first two, Benzene and Ferrocene, are used to assess AEGISS on well-known benchmark molecules. We then consider DOBNA, a prototypical thermally activated delayed fluorescence organic light-emitting diode (TADF-OLED) emitter. Finally, we study two Ruthenium(II)-based complexes of increasing size and complexity, chosen as models for photodynamic therapy (PDT). This latter case study is motivated by the presence of low-lying excited states and strong electronic correlation, for which active space selection is particularly critical. In addition, these systems have recently attracted interest as promising targets for near-term quantum computing applications and have been used in our recent ADAPT-VMPE~\cite{ADAPTVMPE} paper, which highlighted the importance of accurate active-space construction for variational quantum simulations of strongly correlated molecular systems. The results and analysis are presented in Section~\ref{sec:results}, computational details in Section~\ref{sec:computational_details}, and conclusions in Section~\ref{sec:conclusion}. Additional benchmarks and supporting information are provided in the appendix.

\section{Background theory and previous efforts}
\label{sec:background}
In the early 1990s, both CASSCF~\cite{Roos1980,Andersson1990} and CASPT2~\cite{Andersson1992} methods became widely used, with CASPT2 among the first to provide quantitative results for molecular excited states of photochemical interest~\cite{roos08a}. The CASSCF method optimizes a wavefunction over a chosen set of active orbitals, allowing accurate descriptions of strongly correlated systems. CASPT2 refines this wavefunction by adding dynamic correlation effects, leading to better energy predictions. The effectiveness of these methods depends on selecting an appropriate active space, which requires deep chemical intuition and knowledge of the electronic structure. Optimizing and automating this selection remains a challenge due to the need to balance essential electronic correlations with computational feasibility, avoid heuristic metrics, and address near-degenerate states or dynamic correlations.

To extend to active orbital spaces beyond limitations of the \textit{exact} CAS approach, $\approx$~CAS(24 electrons, 24 orbitals)~\cite{vogiatzis2017pushing,hu2024stp-das,gao2024distributed,shayit2025breaking}, DMRG approach~\cite{White1992, White93, Schollwock2005, SCHOLLWOCK2011} constitutes an alternative and versatile option. It is a variational algorithm that optimizes a wavefunction expressed as a Matrix Product State (MPS). The bond dimension of the virtual bonds, that ``connects'' the tensors in the MPS and over which the summation runs to yield the scalar coefficient of the (approximated) FCI tensor, controls, along with other factors~\cite{kell14}, both the accuracy and computational cost. The integration of DMRG algorithms into quantum chemistry codes, such as, for example, PySCF~\cite{pyscf, Sun2018, Sun2020} and OpenMOLCAS~\cite{openmolcas0,openmolcas}, has significantly expanded the applicability of FCI and CAS methods. These advances enable accurate wavefunction calculations for systems above the standard FCI limit, counting approximated CAS solution of sizes that range between 20 to 100 orbitals~\cite{chan2002, dres15, mayi17b, zhai2021, xu2023, Lee2023, Sharma2014, kurashige2013}\ and even extends to fully relativistic Hamiltonian simulations \cite{knec14,batt18,hoye22,hoye23a}. These advancements continue to push the boundaries of quantum chemistry, improving the accuracy and efficiency of electronic structure calculations for increasingly complex molecular systems across the periodic table of elements. 

Yet, the selection of an appropriate active space remains \textit{the} most critical step in multi-configurational calculations, as it directly affects the accuracy and feasibility of electronic structure methods. In the following section, we discuss the main methodologies developed for active space selection and their role in improving computational efficiency and accuracy.

\subsection{Background on active space selection methods}
\label{ssec:as_methods}
In the literature, a collection of different active-space selection methods can be found spanning the last few decades since the advent of multi-configurational approaches in the 1970's. Here, we report the most commonly used ones as of today, especially those from which the protocol presented in this work is inspired by:
\begin{itemize}

    \item \textit{Selection around Fermi-Level}: 
    The simplest approach is to select a set of occupied orbitals below the HOMO, denoted $N_o$, include their associated electrons $N_e$, and choose a desired number of virtual orbitals above the HOMO, $N_v$. This defines a CAS($N_e$, $N_o + N_v$) active space. The underlying assumption is that the most important excitations and correlation effects occur within this energy window, which typically contains the orbitals most relevant to the system’s chemical or physical behavior. However, if important orbitals lie outside this predefined range, a CAS calculation based solely on this selection will inevitably miss essential features of the electronic structure.
    
    \item \textit{Natural Orbital Occupation Number (NOON)}: 
    \teal{This method selects a candidate active space by constructing natural orbitals~\cite{Lowdin1955} and inspecting their occupation numbers~\cite{ROGERS1998, toth2020}. Natural orbitals are obtained as eigenvectors of the one-electron reduced density matrix computed from an approximate correlated wavefunction (commonly MP2). As introduced by L{\"o}wdin~\cite{Lowdin1955}, they satisfy
    \begin{equation}\rho^{\ket{\psi}} = \sum\limits_{{pq}} \rho_{pq}^{\ket{\psi}} = \sum\limits_{pq} \langle\psi | a^{\dagger}_p a_q | \psi\rangle\end{equation} 
    where the eigenvalues of $\rho^{\ket{\psi}}$ are the corresponding NOONs.
    The active space is then defined by selecting natural orbitals with NOONs typically between 0.02 and 1.98 (values indicating partial occupation and hence electronic correlation in $\ket{\psi}$). This strategy is effective because fractional occupations highlight orbitals important for multi-reference character. However, defining active spaces using fixed thresholds become unreliable as soon as frontier orbitals exhibit similar NOONs leading to extremely crowded spaces. Furthermore, NOONs are often derived from MP2 due to its favorable scaling, even though MP2 may be an inadequate reference for certain systems.
    }
    
    \item \textit{ABC scheme}: 
    \teal{The ABC scheme~\cite{bao2018} is an automated procedure for defining active spaces in multi-reference methods such as MS-CASPT2 and MC-PDFT~\cite{Li_Manni2014, Li_Manni2016}, removing the need for manual orbital selection. It relies on three parameters, A, B, and C, which determine the number of doubly occupied orbitals (holes), virtual orbitals (particles), and additional excited states included to ensure balanced treatment of near-degenerate configurations. The protocol involves three calculations: identifying key occupied orbitals, identifying relevant virtual orbitals, and performing a state-averaged CASSCF on the dominant hole/particle set. The resulting multi-configurational wavefunction is then used as the starting point for MS-CASPT2 or MC-PDFT to compute excitation energies.}
    
    \item{\teal{\textit{Active Space Selection based on 1st order perturbation theory (ASS1ST)}: ASS1ST  ~\cite{ass1st2019, ass1st2020}, is a bottom-up automated active space selection scheme designed to remove the system-specific expertise typically required for multi-configuration calculations. Starting from a minimal seed active space, ASS1ST runs strongly contracted-NEVPT2 and diagonalizes the internal and external blocks of the unrelaxed first-order 1-RDM to obtain quasi-natural orbitals and their occupation numbers (NOONs). The actie space is extendend iteratively self consistently by including orbitals with occupation far from 2 and 0.}}
    
    \item \textit{Atomic Valence Active Space (AVAS)}: AVAS is an automated procedure that allows one to build active spaces starting from a list of atomic valence orbitals and a single determinant wavefunction~\cite{Sayfutyarova_2017_AVAS}. Setting out from a (user) pre-defined set of atomic valence orbitals, it includes all  molecular orbitals with a sufficiently large overlap of the predefined set. To this end, AVAS uses a simple linear algebra rotation between the set of occupied and virtual orbitals in order to maximize the desired atomic valence character. In particular, the aim is to isolate the part of the reference wavefunction which involves the set of preselected atomic orbitals from the part which is not related, and define the active space on the first one. 
    
    \item\teal{\textit{Imposed Automatic Selection and Localization of Complete Active Spaces (iCAS)}: iCAS is an automated procedure that constructs an active space from a user-specified list of target atomic valence orbitals (AVAOs) and a single-determinant reference wavefunction~\cite{iCAS}. The preselected AVAO set is first transformed into a set of pre-localized molecular orbitals, which are then used as probes to identify the occupied and virtual HF orbitals that maximally match them. Finally, a localization step yields a chemically
    interpretable set of active orbitals spanning the imposed atomic character. In contrast to AVAS, where the active-space dimension emerges from a singular-value threshold on the projection, in iCAS the number of selected occupied and virtual active orbitals is fixed a priori by the number of occupied and virtual AVAOs in the target set, and no cutoff parameter is required.}
    
    \item \textit{Single-orbital entropy (SOE)}:
    \teal{The single-orbital entropy is a quantum‑information metric widely used in quantum chemistry to analyze electron correlation and guide active space selection. It forms the core of the automated procedure \textit{AutoCAS}.\cite{Stein2016, Stein2017, Stein2019, Bensberg2023}
    Physically, the single‑orbital entropy $S(1)_p$ measures how strongly a spatial orbital $p$ is entangled with the rest of the system.}
    \begin{figure}[!htb]
    \centering
    \includegraphics[scale = 0.55, trim={1.8cm 4.2cm 1.7cm 4.2cm}, clip]{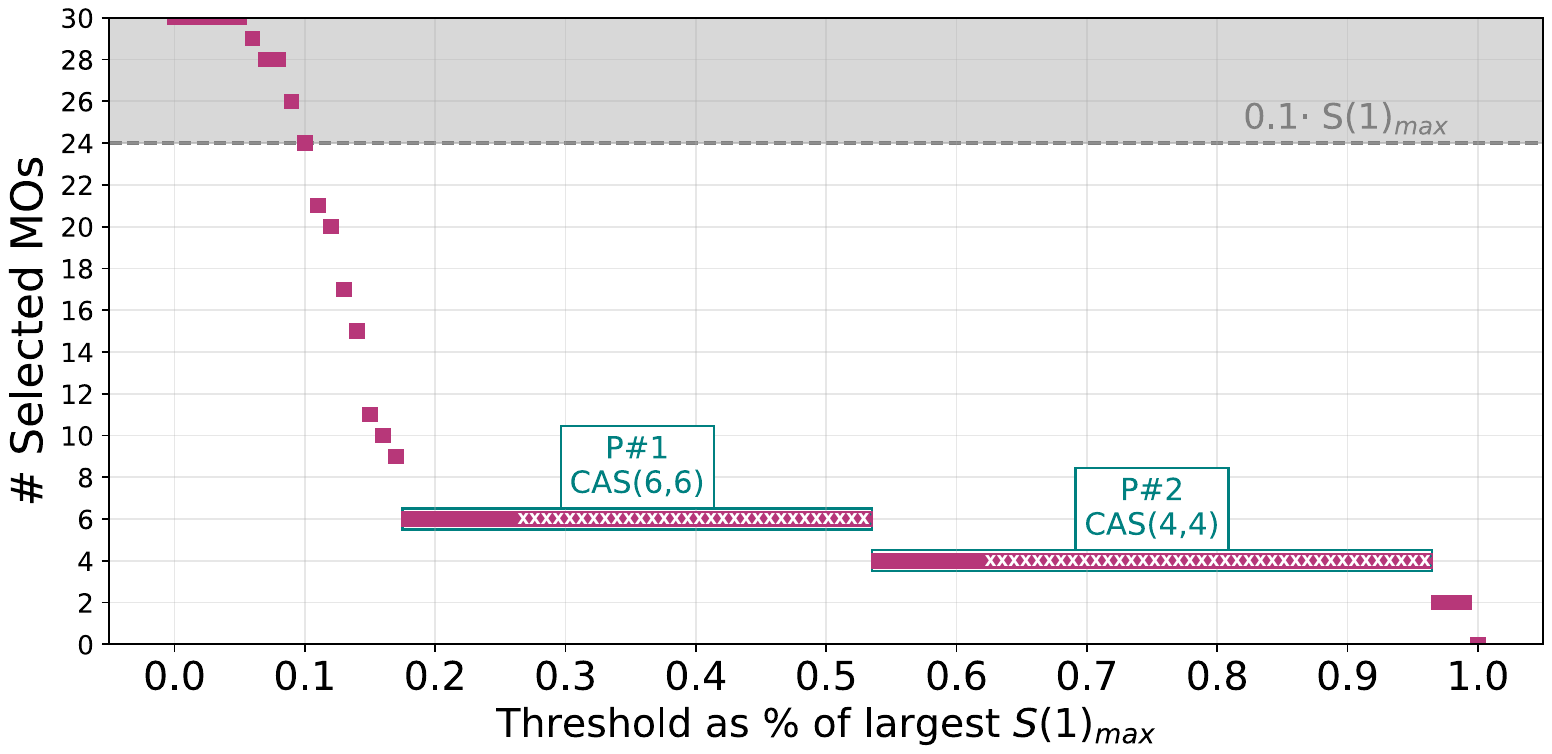}
    \caption{\teal{Examples of threshold plot for a C$_6$H$_6$ molecule in cc-pVTZ basis. The single orbital entropies have been extracted from a DMRG(30,30) calculation with a maximum bond dimension of 200. Each $\blacksquare$ corresponds to how many MOs have an S(1)$_p$ above the fixed fraction. The shaded area shows a 10\% threshold.  In this case, 24 orbitals have an entropy above this  threshold. The white ''$\boldsymbol{\times}$''s displays when a plateau is identified. In the present case, two plateaus have been identified and they correspond to a CAS(6,6) and a CAS(4,4), respectively.}}
    \label{fig:th_plot_examples}
    \end{figure}
    \teal{Large values of $S(1)_p$ indicate orbitals that are strongly correlated and therefore good candidates for inclusion in the active space, whereas orbitals with small $S(1)_p$ contribute mainly dynamical correlation and can typically be excluded. AutoCAS uses $S(1)_p$ computed from a partially converged (low‑bond‑dimension) MPS to construct threshold plots, as described in Ref.~\cite{Stein2017}.
    In such plots, the number of orbitals whose entropy exceeds a chosen fraction of the maximum value $S(1)_{max}$ is shown as a function of that threshold. Plateaus in the threshold plot correspond to stable choices of the active space and enable a straightforward automated selection.}

\end{itemize}
\teal{While the list of active space selection approaches available is longer and contains the effort of many other eminent research groups, we, herein, only gave details about the most relevant ones to our work. We direct the interested readers to further publications, e.g.~\cite{king2021, king2023, Guo2021, Schilling2023, asfcomparison}.}

\teal{To recap, active-space selection methods fall broadly into two conceptually distinct families. The first constructs the active space directly from a single-determinant reference wavefunction, typically by projection onto a user-specified set of valence atomic orbitals, AVAS, and the more recent iCAS, are representative examples.
These methods are generally inexpensive and they rely on the chemical intuition of the user to specify the relevant AOs in advance, but cannot themselves identify orbitals carrying significant static correlation. The second family relies on a preliminary correlated calculation, typically MP2, NEVPT2 or a low-bond-dimension DMRG, to provide information about orbital correlation or natural-orbital occupations, which is then used to define the active space (e.g ABC,
ASS1ST and AutoCAS). These methods do not require a chemical guess but incur the cost of the preliminary correlated step. AEGISS combines the two families: a correlated DMRG calculation provides per-orbital entropy information, and a subsequent AVAS-like projection onto user-specified AO labels extracts the chemically meaningful subset.}

\section{Methodology}
\label{ssec:workflow}
\begin{figure}[!htb]
    \centering
    \includegraphics[scale = 0.65]{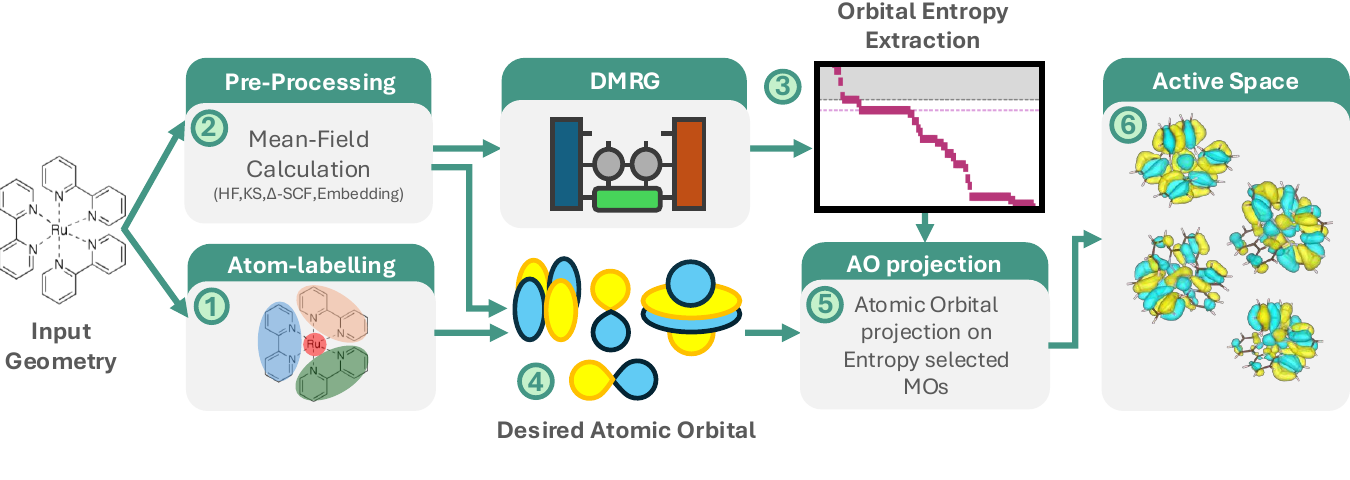}
    \caption{Pictorial representation of the AEGISS workflow. In \textcircled{\small{1}}} the molecule is possibly split into atom clusters, and the atomic orbitals of interest,\textcircled{\small{4}}, are defined according to this subdivision. On the decorated geometry, \textcircled{\small{2}}, builds the initial guess molecular orbitals. An approximate DMRG calculation is then run to pre-select a subset of molecular orbitals, \textcircled{\small{3}}. On this set, the atomic orbital projection is performed, \textcircled{\small{5}}. Finally, \textcircled{\small{6}}, the active space is put together, refining the selection if required.
    \label{fig:workflow}
\end{figure}

\teal{In this work, we propose a new semi-automated protocol for defining active spaces which can produce outputs applicable to both classical active-orbitals-based methods (e.g., CASSCF, CASPT2) and hybrid classical-quantum computing calculations. Our new approach combines the best features of two of the previously mentioned active space selection methods, namely AVAS and AutoCAS, in order to define a \textit{State-average} or \textit{State-Specific} active space selection procedure that takes into account both highly entropic orbitals (i.e., the most correlated ones) as well as chemically and process-specific essential orbitals.} \teal{
AEGISS steps are illustrated in Figure \ref{fig:workflow}.}

\begin{enumerate}[label=\large\protect\textcircled{\small\arabic*}]
    \item \textbf{Atom Labeling.}
The molecule is partitioned into clusters $\mathcal{C}_i$, where each atom $X$ in cluster $i$ is labeled $X_i$. The union of all clusters defines the full molecular set $\{\mathcal{R}\}$. A coarser labeling provides a general description, while a finer one targets specific atomic orbitals. The fragmentation is defined based on where the chemical process of interest occurs. If no prior knowledge about the system is known, this step can be skipped. \teal{General guidelines can be found in Section~\ref{apx:atom-labeling} of the Supplementary Information.}
\begin{figure}
    \centering
    \includegraphics[width=0.3\linewidth]{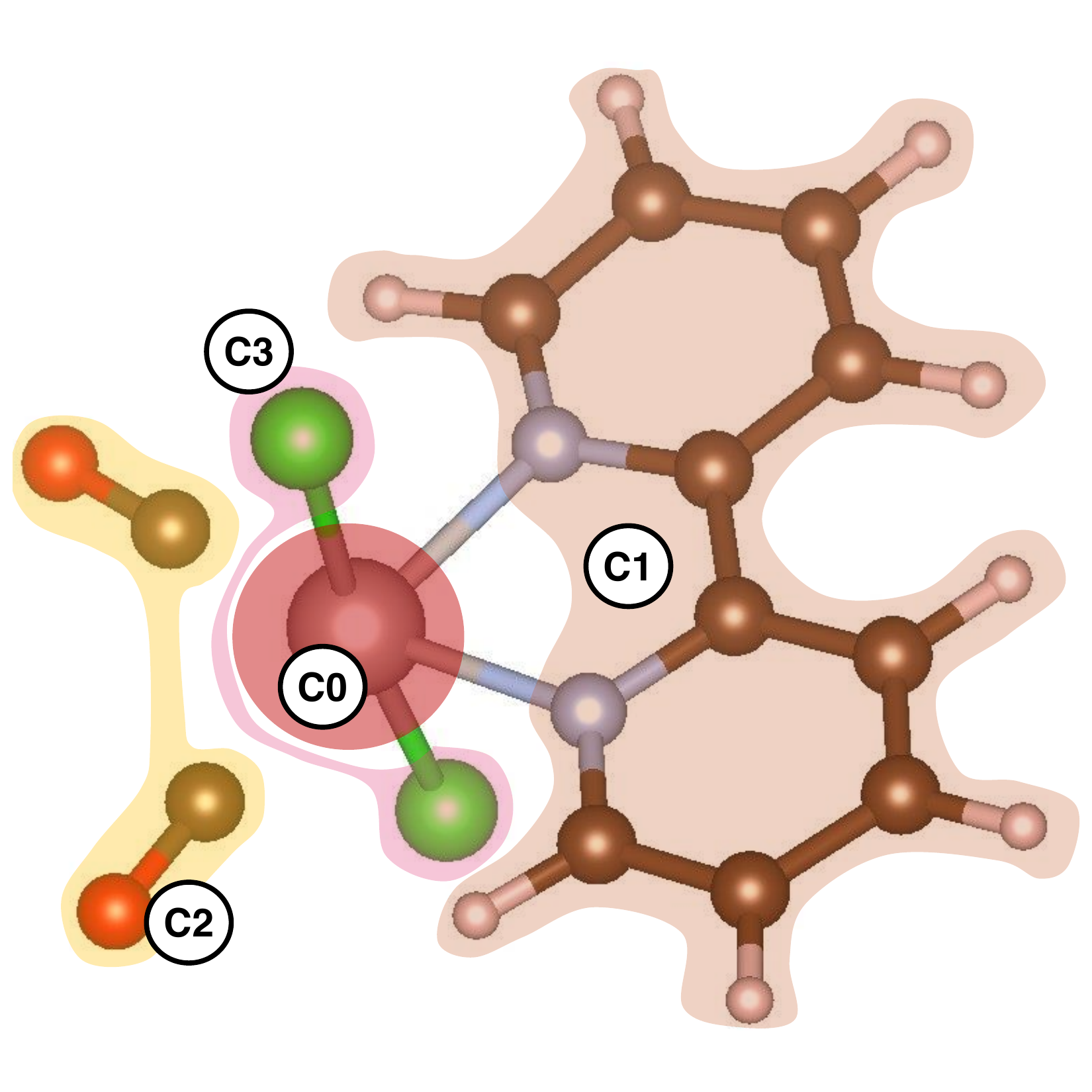}
    \caption{Visual representation of the labeling used for Trans-Cl. Each of the color represent a different fragment, $\mathcal{C}_0$-$\mathcal{C}_3$, of the molecule. In this case, the complex has been split: Ruthenium atom ($\mathcal{C}_0$), bipyridine ($\mathcal{C}_1$), and the two coordinated COs ($\mathcal{C}_2$). \teal{Finally, the two Cl atoms are grouped together ($\mathcal{C}_3$).}}
    \label{fig:transcl_frags}
\end{figure}
    \item \textbf{Mean-Field Calculation.}
A Hartree-Fock (HF)~\cite{aszabo82} (or Kohn-Sham DFT~\cite{KS1965}) SCF calculation yields the MO coefficient matrix $[\mathbf{C}]_{\mu p}$. If needed, it is possible to operate within an embedding framework to account for solvation or other external effects. If the process of interest is localized, one identifies an active region $\mathbf{A}$ and retains only the corresponding orbitals $\{\psi_p^{\mathbf{A}}\}$ with coefficient matrix $[\mathbf{C}^{\mathbf{A}}]_{\mu p}$ of dimension $N_{\text{AO}} \times N_{\mathbf{A}}$. Fragment-based embedding techniques, localized orbitals, or natural orbitals can assist this selection. The environment remains untouched; all subsequent steps operate within $\mathbf{A}$.

\item \textbf{Single-Orbital Entropy Screening.} Next, single orbital entropies are computed from a correlated wavefunction that allows to faithfully identify a relevant large pool of correlated frontier orbitals, reducing the search space for the AO projection step.
\begin{enumerate}
    \item \textbf{Correlated Wavefunction in Active Region.}
    A correlated wavefunction is computed inside $\mathbf{A}$ using the desired method $\mathcal{W}$ (e.g., approximated DMRG):
    \begin{equation}
        \ket{\Psi_{\mathcal{W}}} = \ket{\Phi_{\text{occ}}} \wedge \ket{\psi^{\mathbf{A}}_{\mathcal{W}}} \wedge \ket{\Phi_{\text{virt}}},
    \end{equation}
    and construct one-orbital reduced density matrix (1o-RDM)~\cite{Legeza2003,Rissler2006,bogu15a}
    \begin{equation}
    \rho^{p}_{ii'}
    =
    \sum_{\substack{i,j\in{\{0,\uparrow,\downarrow,2\}}\\\mathbf{n}\neq n^p}}{
    \bra{\psi_{\mathcal{W}}^{\mathbf{A}}}\ket{\mathbf{n}}\ket{i}\bra{j}\bra{\mathbf{n}}\ket{\psi_{\mathcal{W}}^{\mathbf{A}}}},
    \label{eqn:1o-rdm}
    \end{equation}
    where $i,j$ are the occupation of the orbital $p$ and $\ket{\mathbf{n}}$ are the occupation of the other orbitals, and it corresponds to the reduce configurational analogue of one-particle RDM (1-RDM). When access to the full 1o-RDM is impractical, for instance, when only the 1-RDM is available from the correlated solver, or when computing the relevant 2-RDM diagonal is too costly~\cite{Rissler2006,bogu15a}, an approximate scheme for estimating the SOE from natural occupation numbers~\cite{Stein2016,Stein2017MolPhys,Stein2019} is reported in Section~\ref{apx:alt_soe} of the SI.
    
    Here, the choice of the method depends critically on the size of the active region selected, as discussed in Section~\ref{ssec:scalability}.  \teal{In practice we recommend choosing $N_A$ as a Fermi-level window a few times larger than the expected final CAS size, and verifying that the highly entropic orbitals lie well inside the window.}
   \item \textbf{Entropy-based reduction.}
    For each orbital $p$, the SOE~\cite{Legeza2003,Rissler2006,bogu15a} is computed as
   \begin{equation}
       S(1)_p = -\sum_{\alpha=1}^{4} \omega_{\alpha,p} \ln \omega_{\alpha,p},
       \label{eqn:s1}
   \end{equation}
   where $\{\omega_{\alpha,p}\}$ are the eigenvalues of $\rho^p$. All orbitals satisfying $S(1)_p > \tau_E$, with $\tau_E = S(1)_{\max}/10$~\cite{Stein2016}, are retained, yielding the entropy-selected set $\{\psi^{\mathbf{A}}_{E,p}\}$ with coefficient matrix $[\mathbf{C}^{\mathbf{A}}_E]_{\mu p}$ of dimension $N_{\text{AO}} \times N_E$.
   
\end{enumerate}

\item \textbf{Atomic Orbital Selection.}
   Chemically relevant AOs are identified from prior chemical knowledge. For each atomic cluster label $X_i$, the desired AO type ``$\text{AO}_s$'' is specified, forming combined labels ``$X_i\;\text{AO}_s$''. These are grouped into one or more AO sets $D \in \mathcal{D}$; the same AO label may appear in multiple groups.
   
\item \textbf{Atomic Orbital Projection.} After screening the MOs based on their single orbital entropy, we perform the second screening searching for the orbitals that chemically align with the process under study. The contributions of the selected AO-label, defined in the previous step, are then identified using a projection method on atomic orbitals built from a minimal basis-set.

For the AO projection, following the AVAS active-space selection
procedure, an auxiliary minimal atomic orbital basis (MinAO) is
defined, with coefficient matrix $[C^{\mathrm{mAO}}]_{\nu p}$. For
each AO group $D$, we restrict this minimal basis to the AOs
listed in $D$ and compute the overlap of those AOs against the
MOs of the active region $\mathbf{A}$ directly:
\begin{equation}
  [S^D]_{\eta\mu}
  \;=\;
  \int \phi^{\mathrm{mAO}}_{\eta}(\mathbf{r})\,
        \phi^{\mathbf{A}}_{\mu}(\mathbf{r})\,
        \mathrm{d}\mathbf{r},
  \qquad
  \eta \in D,\;
  \mu = 1,\ldots,N_{\mathrm{AO}}.
  \label{eq:overlap-D}
\end{equation}
The resulting matrix has shape $N_D \times N_{\mathrm{AO}}$, where
$N_D$ is the number of AO basis functions in $D$. This is the
projected overlap that enters the subsequent steps; defining it
directly on the AO subset removes the need for an explicit
projection matrix $P^D$.

The residual coefficient matrix of $[S^D]_{\eta\mu}$ on the
entropy-selected active-region MOs $[C^{\mathbf{A}}_E]_{\mu p}$ is
then computed as
\begin{equation}
  [O^D_E]_{\eta p}
  \;=\;
  \sum_{\mu} [S^D]_{\eta\mu}\,[C^{\mathbf{A}}_E]_{\mu p},
  \label{eq:OD}
\end{equation}
with $\eta \in D$, $\mu$ running over the AOs of the original
basis, and $p$ over the entropy-selected MOs. The contribution of
the AOs in $D$ to each candidate MO is finally obtained as
\begin{equation}
  w^D_p
  \;=\;
  \sum_{\eta \in D} [O^D_E]_{\eta p}.
  \label{eq:wDp}
\end{equation}
Figure~5 summarises the procedure of this step.

\item \textbf{Active-Space Construction.}
The union of all selected orbital sets naturally produces the final active space, but this initial selection may require refinement to ensure optimal balance between accuracy and computational feasibility

   The final active space is the union $\{\psi_r\} = \bigcup_{D \in \mathcal{D}} \{\psi_r^D\}$. 
   \teal{
    Given the complexity of active space selection, the AEGISS pipeline is designed to prioritize robustness over precision, adopting a conservative strategy that avoids discarding potentially important orbitals through overly narrow criteria. This is particularly critical as system complexity increases. The selected orbital sets $\{\psi_r^{D}\}$ for each group $D \in \mathcal{D}$ can be further analyzed and refined in terms of size and balance through two complementary approaches: manual screening and automated post-processing.}
\end{enumerate}

The entropy-based screening identifies the most correlated orbitals, while atomic orbital projection ensures chemical relevance. This five-step workflow intentionally adopts a conservative selection strategy that minimizes the risk of excluding chemically important orbitals. The resulting candidate active spaces can then be refined where necessary for particularly challenging systems. This optional refinement allow users to balance computational cost with chemical accuracy based on their specific research goals, making AEGISS adaptable to diverse molecular systems and research objectives.

\subsection{Scalability}
\label{ssec:scalability}
\teal{\noindent A central motivation for this work is to control computational cost without compromising accuracy, while ensuring scalability to large systems such as biomolecules. In this respect, existing approaches such as AutoCAS can become prohibitively expensive and difficult to scale, limiting their applicability in these regimes.
The computational cost of our procedure is partly dominated by the initial mean-field step (HF or KS-DFT), which scales as $\mathcal{O}(N^3)$ and $\mathcal{O}(N^4)$, respectively, with $N$ the system size—a cost shared with other active space selection methods.} \teal{However, as the size of the active region $\mathbf{A}$ increases, the extraction of single-orbital entropies becomes the main bottleneck. The correlated wavefunction method $\mathcal{W}$ employed within  $\mathbf{A}$ scales as $\mathcal{O}(f(|\mathbf{A}|))$, where $|\mathbf{A}|$  denotes the number of active orbitals and associated electrons, and $f$ depends  on the choice of method—for example, $\mathcal{O}(|\mathbf{A}|^6)$ for CCSD or  exponential scaling for exact diagonalization. As $|\mathbf{A}|$ grows, the cost  of computing single-orbital entropies from the correlated wavefunction rapidly dominates.} \teal{In contrast, the AO-projection step—dominated by the computation of the cross-basis overlap matrix $\mathcal{O}(N_{\mathrm{mAO}} N_{AO})$ and the subsequent MO projection $\mathcal{O}(N_D N_{AO} N_E)$—remains negligible. A detailed scaling analysis is provided in the Supplementary Information (Section B).
The overall cost can be effectively contained and the scalability tuned by controlling these key factors—namely the size of the active space $k$ and the DMRG bond dimension $m$ governing the entropy evaluation, together with the extent of the projection space—providing practical handles to balance accuracy and efficiency in large-scale applications.}

\section{Results}
\teal{To demonstrate that AEGISS reliably recovers active spaces across different correlation regimes, we first apply it to two universally adopted benchmark systems, benzene and ferrocene. These are deliberately chosen because their canonical active spaces are well established and have served as validation targets for previous active-space selection schemes.
Having established its behavior on canonical benchmarks, we next assess the robustness of AEGISS in more chemically realistic settings by considering two important classes of photoactive molecules: thermally activated delayed fluorescence (TADF) emitters and photodynamic therapy (PDT) agents. These systems present considerably more intricate active-space selection problems due to the simultaneous presence of strong correlation, excited-state effects, and chemically extended valence manifolds. Accurate theoretical descriptions of photoactive systems are notoriously challenging because electronically excited states can exhibit qualitatively different characters, including $\pi$-$\pi^*$ excitations and charge-transfer states. Moreover, reliable prediction of excitation energies often requires balancing static and dynamic correlation effects alongside geometric relaxation and environmental contributions\cite{Matsika2018}. TADF emitters and PDT agents therefore provide stringent tests of whether AEGISS can identify chemically meaningful active spaces in regimes that extend well beyond traditional benchmark systems.}

\teal{In addition to generating active spaces, we validate the selected orbital manifolds through subsequent multi-configuration calculations, including CASSCF-based approaches, perturbation theory corrections, and benchmark the resulting excited-state properties against TDDFT calculations. All AEGISS-related and CAS calculations have been run in PySCF~\cite{pyscf, Sun2018, Sun2020}, Block2~\cite{zhai2021,zhai2023block2}, and the \texttt{dmrgscf} interface between the two. Further computational details are reported in Section~\ref{sec:computational_details} in the Supporting Information.}

\label{sec:results}
\subsection{\teal{Benzene: \texorpdfstring{C$_6$H$_6$}{C6H6}}}
\teal{
Benzene is the textbook benchmark for $\pi$-conjugated systems. Its six $\pi$ electrons distributed over six $\pi$ molecular orbitals define the canonical CAS$(6,6)$ active space, and Figure~\ref{fig:th_plot_examples} in Section~\ref{sec:background} already illustrates how single-orbital entropies recover precisely this space from a DMRG$(30,30)$ calculation, with a secondary plateau at CAS$(4,4)$. For AEGISS, benzene serves as a minimal sanity check that the combination of entropy screening and AO projection, (for the geometry chosen, 2p$_z$ contributions on the carbons) cleanly isolates the $\pi$/$\pi^*$-manifold from the surrounding $\sigma$-framework. This is a prerequisite for any application to larger $\pi$-conjugated systems such as the TADF emitters and the polypyridyl ligands of the Ru(II) complexes.}

\begin{figure}[!hbt]
    \centering
    \includegraphics[width=0.6\linewidth,trim={0 3.5cm 0 2.7cm}, clip]{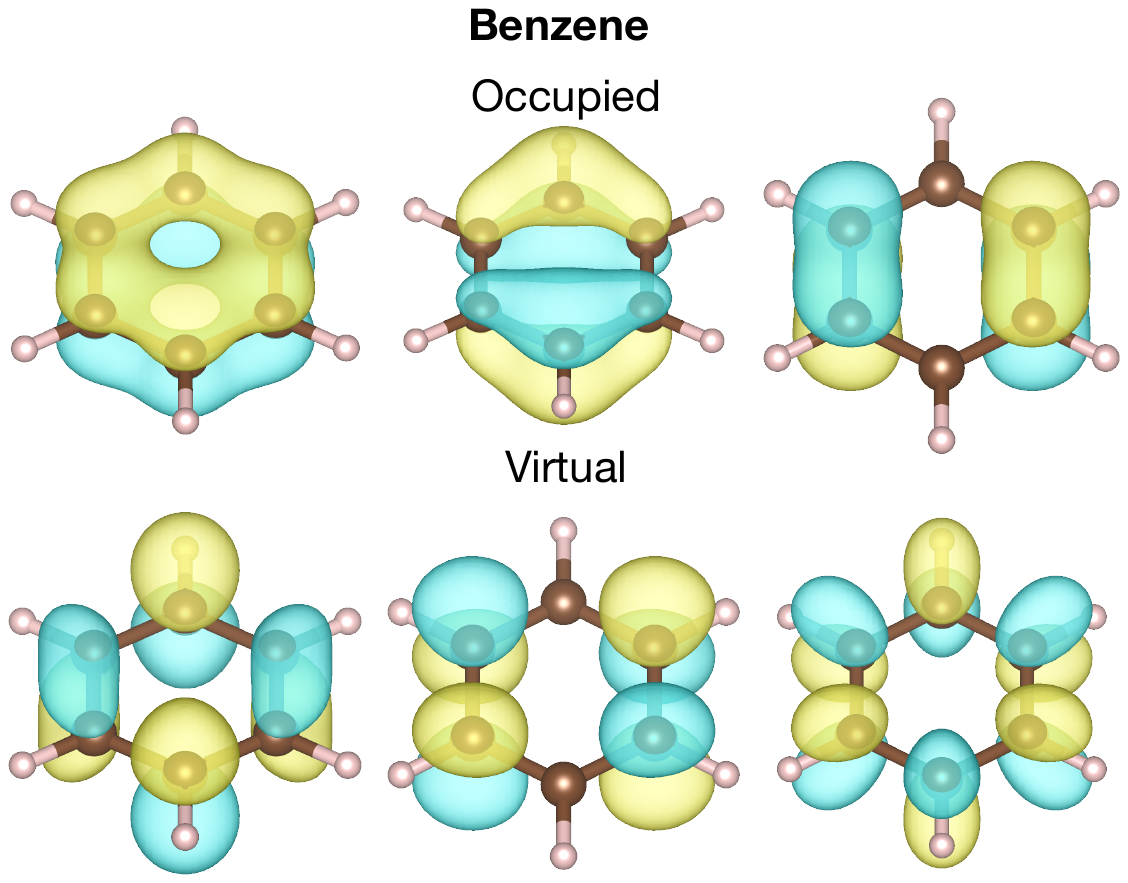}
    \caption{\teal{Final active space identified by AEGISS procedure for the benzene molecule.}}
    \label{fig:benzene_space}
\end{figure}

\begin{table}[!htb]
    \centering
    \begin{tabular}{c|c|c|c|c|c|c|c}
    \toprule
         Guess & Basis & CAS$_{\text{DMRG}}$ &BD& $\tau_E$ & $D$ & $\epsilon_D$ & CAS$_{\text{A}}$   \\\hline\hline
         Canonical RHF& cc-pVDZ & (30e,30o) & 500 & 0.2 & \texttt{"C 2pz"} & 0.5 & (6e,6o)\\ 
         \bottomrule
    \end{tabular}
    \caption{\teal{AEGISS settings for the generating the benzene active-space. CAS$_{\text{DMRG}}$ corresponds to the $N_{\mathbf{A}}$ = 30, selected around the Fermi-level, while $\tau_E$ refers to 20\% of the $S(1)_{max}$, i.e. the maximum SOE amongst all the orbitals in the active subset $\mathbf{A}$. Interested in the \texttt{"C 2pz"} contributions above $\epsilon_{D}$=0.5. Finally, CAS$_{\text{A}}$ is the obtained active space size. }}
    \label{tab:sel_benzene}
\end{table}
\teal{A fixed $D_{6h}$-symmetric geometry was employed for benzene, with standard C–C and C–H bond lengths of 1.397{\AA} and 1.093{\AA}, respectively, laying in the (xy)-plane. Following the AEGISS workflow, defined in Section~\ref{ssec:workflow}, we perform a restricted Hartree–Fock (RHF) calculation to build the initial guess molecular orbitals (MOs), defining an initial orbital space of 116 MOs. For the entropy extraction, we perform a DMRG calculation on a CAS(30e,30o), with a $N_\textbf{A}$=30. We restricted on smaller subset of active orbitals in order to make the entropy screening manageable. The selection of this subset of orbitals is completely arbitrary and in this example canonical RHF are used. From PySCF, the initial FCIDUMP is created and from it, we performed a spin-restricted (SU2) DMRG calculation with bond dimension 500 and then converted to non-spin adapted (SZ) MPS. From the approximated wavefunction, using Equation~\ref{eqn:s1} we compute the SOE, $S(1)_i$, associated with each orbital $\psi_i$ in the active subset, $\{\psi_j\}_{j\in \textbf{A}}$. Applying a 20\% cutoff, $\tau_E$ = 0.2, on the relative entropies, we obtain a reduced set of entropy-screened MOs, $\{\psi^{\mathbf{A}}_{E,p}\}$, composed of 11 orbitals. In the second screening step, the AO-projection, we identify the projection-weights, $w_p^{D}$, for each previously screened orbital, $\psi_p$, relative to the AO-label $D$ = \texttt{"C 2pz"}. We then select all the entropy-screened orbitals $\psi_p$ which projection-weight, $w_p^D$, associated with the AO-label $D$, is above the threshold 0.5. After this second selection, the resulting set of orbitals, $\{\psi_r^{D}\}$, is composed of 6 orbitals, 3 occupied and 3 virtual, defining therefore a CAS(6e,6o). Table~\ref{tab:sel_benzene} sums up the setting for selection through AEGISS.}
\teal{The orbital compositions confirm that AEGISS selects exactly the six $\pi$-network orbitals of benzene. The entropy- and AO-projection-based selection criteria collectively recover the canonical (6e, 6o) active space. The active space is reported in Figure~\ref{fig:benzene_space}}

\teal{For validation, we compare CASCI and CASSCF of the benzene ground-state energy with the active space found by AEGISS with the one generated by AVAS. In AVAS, the space is identified targeting the \texttt{2pz} AOs of the carbon (\texttt{C}) atoms with a threshold of 0.1 (which is the default threshold in PySCF implementation of the selection procedure). Table \ref{tab:benzene_res} reports the CASCI and CASSCF ground-state energies for benzene obtained with active spaces selected by AEGISS and AVAS, both yielding a CAS(6e,6o) active space corresponding to the six $\pi$-network orbitals. The two methods differ in their CASCI energies, with AVAS recovering a larger fraction of the correlation energy at the CASCI level (E$_\text{corr}$ = -0.069 Ha) compared to AEGISS (E$_\text{corr}$ = -0.054 Ha). This is expected, as AVAS applies an explicit projection onto the target AO subspace, rotating the molecular orbitals to maximize their $\pi$-character prior to the multi-configurational calculation, whereas AEGISS operates directly on the canonical RHF orbitals without any pre-rotation. Despite this difference, both methods converge to identical CASSCF energies ($E$ = -230.793770 Ha, E$_\text{corr}$ = -0.072 Ha), demonstrating that the active space selected by AEGISS spans the same orbital space as AVAS upon orbital optimization. This agreement validates the entropy- and AO-projection-based selection of AEGISS, confirming that the correct chemically relevant degrees of freedom are identified.}
\begin{table}[!htb]
    \centering
    \begin{tabular}{c|c|c|c|c}
    \toprule
    Method & CASCI & E$_{corr}$ & CASSCF & E$_{corr}$\\\hline\hline
    AEGISS & -230.776& -0.054 & -230.794 & -0.072    \\
    AVAS  &-230.791 &-0.069 & -230.794 & -0.072 \\
    \bottomrule
    \end{tabular}
    \caption{\teal{Comparison between AEGISS and AVAS active spaces for multi-configurational calculations, CASCI and CASSCF, ground-state energy for benzene molecule. Energy CASCI and CASSCF reported in Hartree (Ha), as well as the correlation energy w.r.t. to the HF solution, E$_{corr}$.}}
    \label{tab:benzene_res}
\end{table}

\subsection{\teal{Ferrocene:\texorpdfstring{Fe(C$_5$H$_5$)$_2$}{Fe(C5H5)2}}}
\teal{Ferrocene complements benzene by probing transition-metal correlation in a structurally simple, high-symmetry setting. It presents a sandwich structure that couples the Fe 3d shell to two cyclopentadienyl $\pi$-systems. Ferrocene is a standard test case for CASSCF, DMRG, and selected-CI selection schemes precisely because naive frontier-orbital choices fail: the orbitals to be included are not simply those nearest the Fermi level, but those identified jointly by their entanglement and by their atomic character. This makes it an ideal stress test for the AEGISS workflow, which is designed to combine exactly these two criteria.}

\begin{figure}[!hbt]
    \centering
    \includegraphics[width=0.7\linewidth, trim={0 2.2cm 0 2.7cm}, clip]{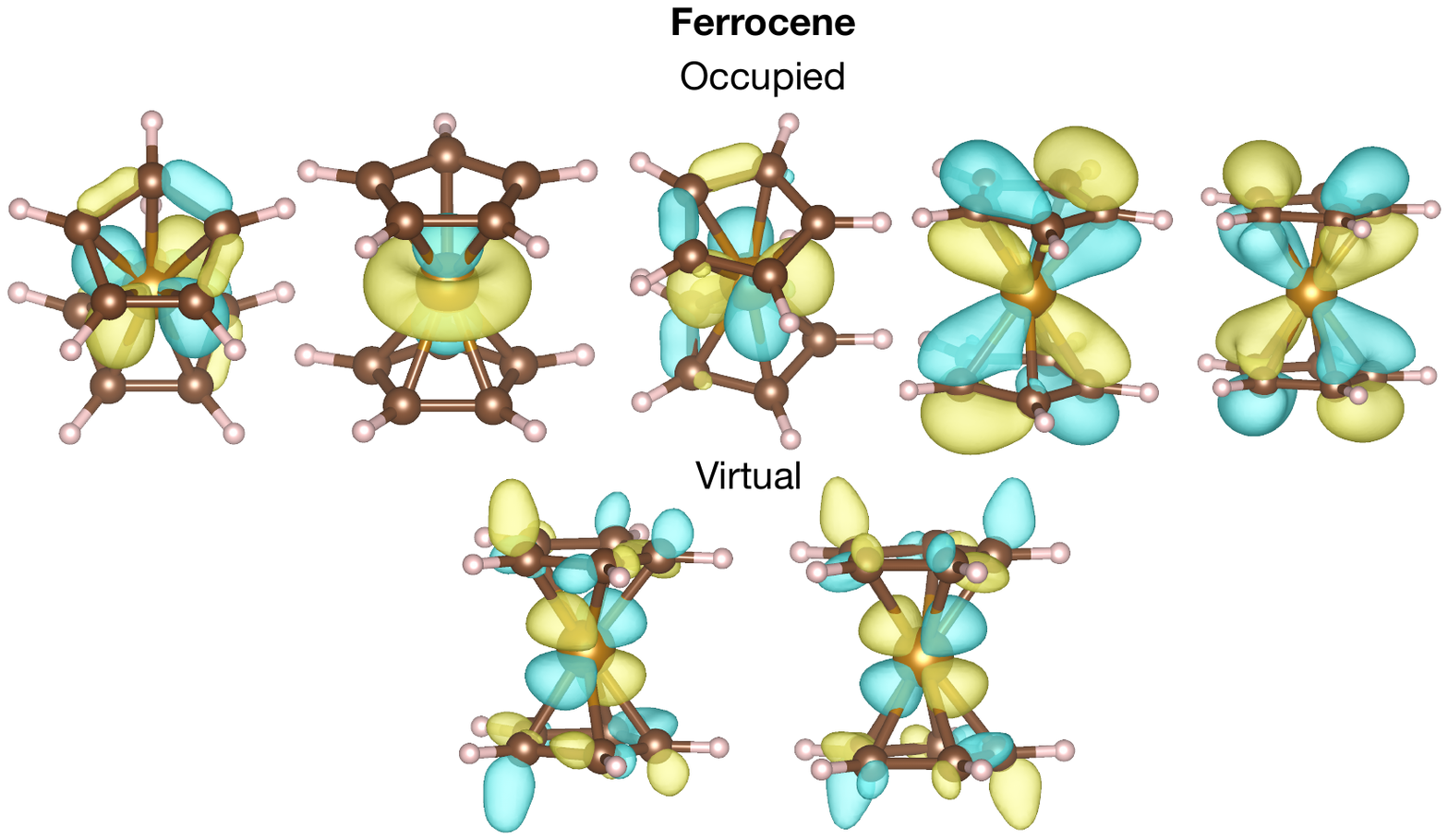}
    \caption{\teal{Occupied and virtual final active space identified by AEGISS procedure. It is possible to identify the pure 3d contributions, in both occupied (first three orbital in the 1st row) and virtual space, as well as the two occupied hybrid Cp 2p$_z$-Fe 3d (last two orbital in the 1st row).}}
    \label{fig:ferrocene_space}
\end{figure}

\begin{table}[!htb]
    \centering
    \begin{tabular}{c|c|c|c|c|c|c|c}
    \toprule
         Guess & Basis & CAS$_{\text{DMRG}}$ &BD& $\tau_E$ & $D$ & $\epsilon_D$ & CAS$_{\text{A}}$  \\\hline\hline
         \makecell{Canonical RHF \\ X2C~\cite{x2c_1, x2c_2}}& cc-pVTZ-DK & (40e,40o) & 500 & 0.1 & \texttt{Fe 3d} & 0.5 & (10e,7o)\\ 
         \bottomrule
    \end{tabular}
    \caption{\teal{AEGISS settings for the generating the ferrocene active-space. The initial guess is obtained with RHF taking into account also scalar relativistic corrections (X2C). CAS$_{\text{DMRG}}$ corresponds to the $N_{\mathbf{A}}$ = 40, selected around the Fermi-level, while $\tau_E$ refers to 10\% of the $S(1)_{max}$, i.e. the maximum SOE amongst all the orbitals in the active subset $\mathbf{A}$. Interested in the \texttt{Fe 3d} contributions above $\epsilon_{D}$=0.5. Finally, CAS$_{\text{A}}$ is the obtained active space size. }}
    \label{tab:sel_ferrocene}
\end{table} 
\teal{We applied the AEGISS procedure, explained in Section~\ref{ssec:workflow}, on ferrocene with a setup as reported in Table~\ref{tab:sel_ferrocene}, while the computational details are collected in Section~\ref{apx:setup_ferrocene}.}
\teal{The identified set, CAS$_{\text{A}}$, is composed of 7 MOs, 5 occupied and 2 virtual orbitals, defining a CAS(10e,7o), as in the reference AVAS paper~\cite{Sayfutyarova_2017_AVAS}. Analyzing in detail the composition of the orbitals, we can notice that 5 find exact correspondence with the Fe 3d AOs, in particular, 3 occupied and the 2 virtual orbitals.  The remaining two occupied ones, present a mixed Fe $\mathrm{3d}_{xz}$ and $\mathrm{3d}_{yz}$ with Cp $\mathrm{2p}_z$, composition shared with the two virtual orbitals, identifying in them the responsible for the bonding with the Cp rings. The space is shown in Figure~\ref{fig:ferrocene_space}.}

\teal{In order to asses the quality of the active space obtained with AEGISS, we performed CASSCF and NEVPT2 on top. In particular, we are interested in the first 7 singlets and 6 triplet states. Therefore, a state-averaged multi-configuration calculation with mixed multiplicity is required. Additionally, we report AVAS results (CASSCF and NEVPT2) from the original method paper~\cite{Sayfutyarova_2017_AVAS}, TDDFT excitation energies computed following Ref.~\cite{Fromager2013} using the PBE functional and extended to include six triplet states, as well as available experimental data~\cite{Ferrocene67, metallocenes71}}.

\begin{table}[!htb]
    \centering
    \begin{tabular}{c|cc|cc|c|c}
    \toprule
         State & \multicolumn{2}{c|}{AEGISS (10e,7o)} & \multicolumn{2}{c|}{AVAS (10e,7o)~\cite{Sayfutyarova_2017_AVAS}} & PBE & Exp. \\
         & CASSCF & (+NEVPT2) & CASSCF & (+NEVPT2) & & \\\hline\hline
 1$^1$E$_2''$ & 2.098 & 2.869 & 2.10 & 2.79 & 2.90~\cite{Fromager2013} & 2.8~\cite{Ferrocene67}, 2.7~\cite{metallocenes71}\\
1$^1$E$_1''$ & 2.169 & 2.786 & 2.17 & 2.87 & 3.03~\cite{Fromager2013} & 2.81~\cite{Ferrocene67}, 2.98\\
2$^1$E$_1''$ & 4.068 & 3.986 & 4.07 & 3.99 & 3.60~\cite{Fromager2013} & 3.82~\cite{Ferrocene67,metallocenes71}\\
\hline
1$^3$E$_1''$ & 0.973 & 1.880 & 0.97 & 1.88 & $\sim$ & 1.74~\cite{Ferrocene67}\\
1$^3$E$_2''$ & 1.070 & 2.031 & 1.07 & 2.03 & $\sim$ & 2.05~\cite{Ferrocene67}\\
2$^3$E$_1''$ & 2.246 & 2.503 & 2.25 & 2.50 & $\sim$ & 2.29--2.34~\cite{Ferrocene67}\\
\bottomrule
    \end{tabular}
    \caption{\teal{Vertical excitation energies of ferrocene (eV) from state-averaged CASSCF and NEVPT2 over a (10e,7o) active space constructed by AEGISS, compared against the AVAS reference values from Ref.~\cite{Sayfutyarova_2017_AVAS}. SA-CASSCF was performed in C$_{2v}$ symmetry with irrep-resolved FCI solvers; each D$_{5h}$ E$_n''$ state corresponds to a degenerate (A$_2$, B$_2$) pair. PBE/TD-DFT and experimental values are reported for comparison.}}
    \label{tab:ferrocene_res}
\end{table}
\teal{Table~\ref{tab:ferrocene_res} collects all the results for multi-configurational methods CASSCF with NEVPT2 correction on top, as well as TDDFT, and experimental data.}

\subsection{\teal{TADF molecule}}
\teal{Organic light-emitting diodes (OLEDs) based on thermally activated delayed fluorescence (TADF) represent a promising new generation of metal-free emitters for high-efficiency optoelectronic devices~\cite{review_TADF}. Efficient reverse intersystem crossing (RISC) requires the singlet-triplet energy gap, $\Delta E_{ST}$, to be comparable to or smaller than $k_B T$, motivating molecular designs with pronounced charge-transfer character. In particular, multiple-resonance (MR) TADF emitters exploit the opposing resonance effects of boron and nitrogen/oxygen atoms to generate short-range charge-transfer excited states characterized by both small $\Delta E_{ST}$ values and narrow emission bands. The interplay between local and charge-transfer excitations, combined with extended $\pi$-conjugation, makes these systems particularly challenging for multi-configurational electronic-structure methods and highly sensitive to the choice of active space.}

\teal{In this work, we focus on DOBNA~\cite{Hatakeyama2016DOBNA, Kim2023DOBNAhost, review_TADF, DOBNA_geom} (5,9-dioxa-13b-boranaphtho[3,2,1-de]anthracene), a prototypical MR-TADF emitter. As the structurally minimal boron/oxygen scaffold within this family, DOBNA exhibits a singlet-triplet gap of approximately 0.15 eV and ultraviolet emission, making it a compact yet representative benchmark system.}
\begin{table}[!htb]
    \centering
    \begin{tabular}{c|c|c|c|c|c|c|c|c}
    \toprule
         Guess & Basis & CAS$_{\text{DMRG}}$ &BD& $\tau_E$ & $D$ & $\epsilon_D$ & CAS$_{A}$ & Name \\\hline\hline
         Canonical RHF & \multirow{5}{*}{6-31G(d)} & \multirow{5}{*}{(40e,40o)} & \multirow{5}{*}{500} & \makecell{0.1\\0.9} & \multirow{5}{*}{\texttt{"[C/B] 2px"}} & \multirow{5}{*}{[0.5]} & \makecell{(18e,18o)\\(10e,10o)}&\makecell{I\\II}\\ \cline{1-1}\cline{5-5}\cline{8-9}
         
         MP2-NOs &  &  &  & \makecell{0.1\\0.9} &  & & \makecell{(18e,18o)\\(10e,10o)} &\makecell{III\\IV}\\ \cline{1-1}\cline{5-5}\cline{8-9}
         
         MP2-(F)NOs &  &  &  & \makecell{0.1\\0.9} &  & & \makecell{(18e,20o)\\(8e,9o)} &\makecell{V\\VI}\\  
         \bottomrule
    \end{tabular}
    \caption{\teal{AEGISS settings for the generating the DOBNA active-space. The initial guess orbitals are Canonical RHF and NOs obtained from a correlated MP2 calculation, starting the previous orbitals. CAS$_{\text{DMRG}}$ corresponds to the $N_{\mathbf{A}}$ = 40, selected around the Fermi-level, while $\tau_E$ refers to 10\%, for spaces I,III, and V, and 90\%, for space II, IV, and VI, of the $S(1)_{max}$, i.e. the maximum SOE amongst all the orbitals in the active subset $\mathbf{A}$. Interested in the \texttt{2px} contributions above $\epsilon_{D}$=0.5 for both carbon (\texttt{C}) and boron (\texttt{B}) atoms. Finally, CAS$_{\text{A}}$ is the obtained active space size.}}
    \label{tab:sel_dobna}
\end{table} 
\teal{For this molecule, we adopted a slightly modified protocol compared to the previous case studies. Preliminary RHF calculations with the 6-31G(d) basis set revealed significant Rydberg-like character already within the first ten virtual orbitals. To account for this feature, orbital selection was additionally performed using MP2 natural orbitals (NOs)~\cite{MP2-NO} as the initial molecular-orbital guess. The workflow as described in Section~\ref{ssec:workflow} was then applied using the settings reported in Table~\ref{tab:sel_dobna}, while computational details are provided in Section~\ref{apx:setup_dobna}.}
\begin{figure}[!hbt]
    \centering
    \includegraphics[width=0.75\linewidth,trim={0 4.5cm 0 2.7cm}, clip]{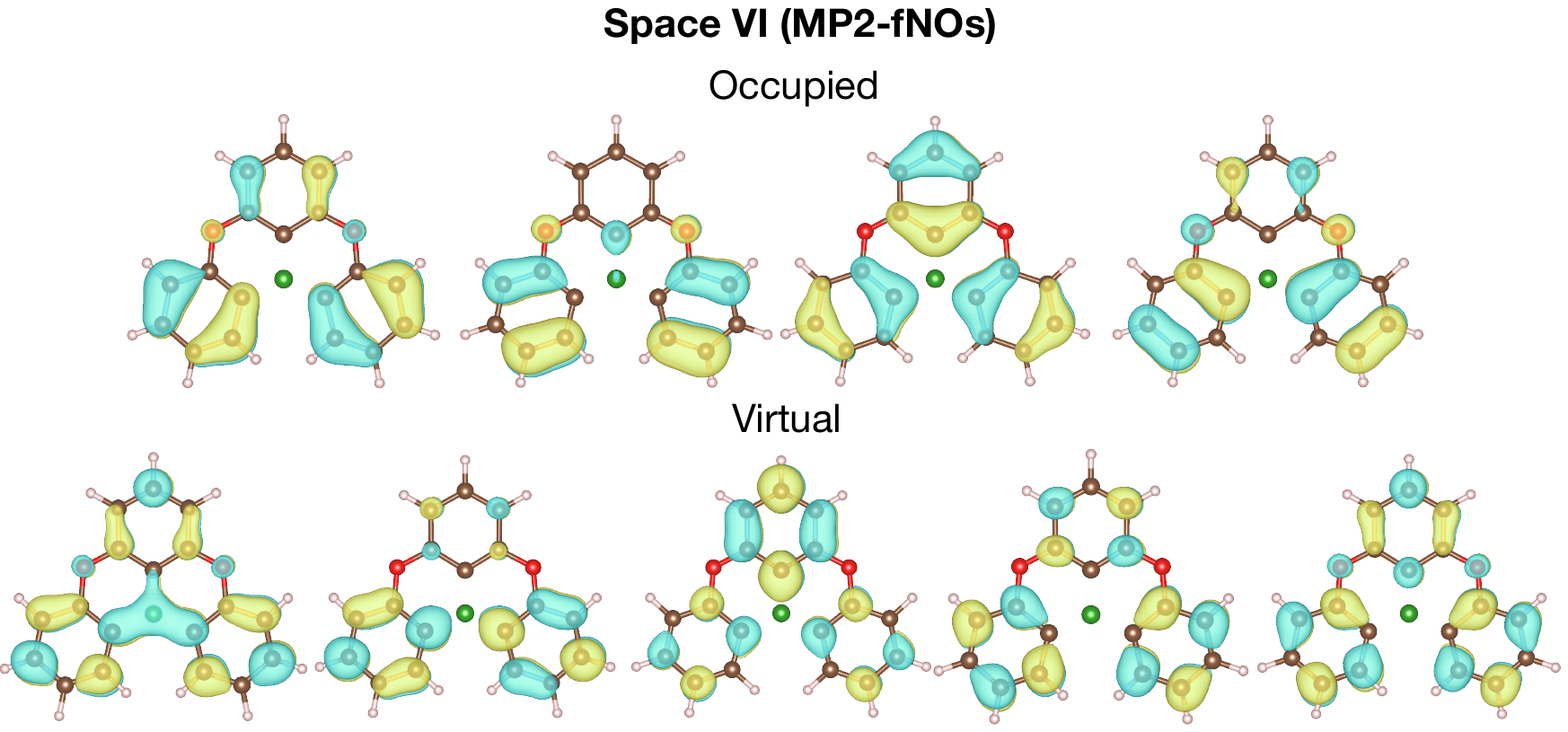}
    \caption{\teal{Occupied and virtual Space (VI) identified by AEGISS procedure on DOBNA using MP2-(F)NOs initial guess and tighter entropy thresholding.}}
    \label{fig:dobna_space_fNOs}
\end{figure}
\teal{AEGISS generated six candidate active spaces. Spaces (I) and (II) were obtained using canonical RHF orbitals with loose and stringent entropy thresholds, respectively. Spaces (III) and (IV) employed MP2 natural orbitals as the initial guess, again using two different entropy cutoffs. Finally, spaces (V) and (VI) were generated from MP2 frozen natural orbitals [(F)NOs]~\cite{SosaFNOs, taube2005frozen, taube2008, verma2021, MP2-NO}, corresponding to loose and stringent entropy selections, respectively.
Comparison of the entropy distributions obtained from RHF and MP2-NO initial guesses reveals markedly different behaviors. In the RHF case, more than half of the orbitals exhibit relative SOE values below 0.1 and are therefore removed during entropy screening. In contrast, the partial correlation already incorporated in the MP2-NO orbitals results in all relative SOEs remaining above this threshold, preventing any orbital removal at this stage. Nevertheless, both distributions display a distinct subset of highly entropic orbitals with SOE values exceeding 90\% of the maximum entropy. The smallest active spaces, (II), (IV), and (VI), are reported in Figure~\ref{sfig:dobna_space_can},~\ref{sfig:dobna_space_NOs}, and ~\ref{fig:dobna_space_fNOs}, respectively.}

\teal{For the loose entropy threshold, both RHF and MP2-NO selections produce a CAS(18e,18o), corresponding to spaces (I) and (III), whereas the more stringent threshold yields CAS(10e,10o) active spaces, namely (II) and (IV). In particular, for space (III), the reduction in active-space size originates almost entirely from the AO-projection step, since entropy screening does not exclude any orbitals. Inspection of the final orbital sets selected by AEGISS shows that the RHF- and MP2-NO-based spaces contain similar orbitals. Given this substantial overlap, subsequent validation calculations show similar results. A different behavior emerges when MP2-(F)NOs are employed. In this case, the resulting active spaces differ substantially from the RHF- and MP2-NO-derived selections. The loose threshold leads to a larger active space (V), whereas the stricter threshold produces a smaller one (VI). Furthermore, both spaces exhibit a less balanced occupied-virtual partitioning, with a noticeably larger fraction of virtual orbitals included in the final selection.}

\teal{We compared the AEGISS-derived active spaces with those constructed via AVAS. For DOBNA, we used the same set of AO labels employed in AEGISS (reported in Table~\ref{tab:sel_dobna}) and tested two cutoff thresholds (see Ref.~\cite{Sayfutyarova_2017_AVAS} for a detailed description of the AVAS thresholding procedure), namely the default value of 0.2 and a tighter value of 0.5. In both cases AVAS recovers the expected $\pi/\pi^*$-manifold, yielding a CAS(18,19). Reducing the space further requires pushing the threshold above 0.80, since all orbitals selected up to that point present singular values above 0.865.}

\begin{table}[!htb]
\centering
\small
\begin{tabular}{c|ccc|ccc|ccc|c}
\toprule
& \multicolumn{3}{c|}{CASSCF} & \multicolumn{3}{c|}{NEVPT2}& PBE & B3LYP &Cam-B3LYP& Exp.\\
Space & $S_1$ & $T_1$ & $\Delta E_{\text{S-T}}$ & $S_1$ & $T_1$ & $\Delta E_{\text{S-T}}$ &\multicolumn{3}{c|}{$\Delta E_{\text{S-T}}$} & ~\cite{gap_dobna,review_TADF} \\\hline\hline
II & 4.663 & 3.664 & 0.999 & 4.939 & 4.208 & 0.731 &~&~&~&~ \\
IV & 4.663 & 3.664 & 0.999 & 4.938 & 4.207 &0.731&\multirow{1}{*}{0.393}&\multirow{1}{*}{0.524}&\multirow{1}{*}{0.743}&\multirow{1}{*}{0.15}\\
VI & 5.352 & 4.299 & 1.053 & 4.560 & 4.117 & 0.443&~&~&~&~ \\
\bottomrule
\end{tabular}
\caption{\teal{DOBNA vertical excitation energies and singlet-triplet gap for the three SA-CASSCF and NEVPT2 calculations. State-averaged on 6 singlets and 5 triplets for the smallest space. Comparison with TD-DFT with three functionals: PBE, B3LYP, and Cam-B3LYP.  All energies in eV. For TD-DFT only the $\Delta E_{\text{S-T}}$ are reported, while the full excitation spectrum is presented in Section~\ref{apx:dobna_benchmarks}. Experimental $\Delta E_{\text{S-T}}$ reported in last column.}}
\label{tab:dobna_three_spaces}
\end{table}

\teal{To assess the quality of the selected active spaces, we performed state-averaged CASSCF calculations followed by NEVPT2 corrections, collected in Table~\ref{tab:dobna_three_spaces}. For spaces (I) and (III), whose size exceeds the practical limits of conventional CASSCF, DMRG-SCF calculations were employed instead. The state-averaged wave functions included the lowest six singlet and five triplet states, resulting in an 11-root mixed-multiplicity optimization.
Comparison of the SA-CASSCF and NEVPT2 results for the two smallest active spaces, (IV) and (VI), shows that the perturbative correction systematically reduces the singlet-triplet gap relative to the CASSCF reference. The MP2-NO-based space (IV) yields a NEVPT2 value of $\Delta E_{\mathrm{S-T}} = 0.73$ eV, whereas the MP2-(F)NO-based space (VI) lowers the gap to 0.44 eV despite containing a smaller occupied subspace. In the Supplementary information the solution obtained by spaces (I), (III) and (V) can be found. For those spaces, given the cost of running DMRG-SCF and PT2 correction on top  on 11 roots, we decided to reduce to 4 roots (2 singlets and 2 triplets). In Table~\ref{tab:dobna_6_spaces}, we report the comparison between CASSCF/DMRG-SCF and NEVPT2 energies for all the spaces with 4 roots. }

\teal{For comparison, TD-DFT calculations were also performed using three exchange-correlation functionals: PBE, B3LYP, and CAM-B3LYP. Experimental estimates of the singlet-triplet gap are available for this system~\cite{review_TADF}. The TD-DFT values span a broad range, from 0.39 eV with PBE to 0.74 eV with CAM-B3LYP, while B3LYP predicts an intermediate value of 0.52 eV, in closer agreement with the NEVPT2 result obtained for space (VI).
Relative to the experimental estimate measured in a PMMA matrix ($\sim$0.15 eV~\cite{review_TADF}), all theoretical approaches considered here, including NEVPT2, overestimate $\Delta E_{\mathrm{S-T}}$. Among the active spaces investigated, space (VI) provides the closest agreement with experiment. The remaining discrepancy is consistent with previous studies of MR-TADF emitters, where accurate reproduction of experimental gaps generally requires an exceptionally balanced treatment of dynamical correlation beyond that captured within the active spaces considered in the present work.}

\subsection{Ruthenium(II)-based complexes for photo-dynamic therapy}
In PDT, light activates a photosensitizer, triggering ISC and energy transfer to surrounding oxygen to generate reactive oxygen species (ROS) that induce cancer cell death. A small singlet–triplet gap ($\Delta$S–T) enhances ISC and reverse ISC, improving ROS production and therapy efficiency \cite{McFarland2019}. Understanding these molecular processes supports advances in non-invasive cancer treatments.
\begin{figure}[!htb]
\centering
    \begin{subfigure}[b]{.20\linewidth}
    \includegraphics[width=\linewidth]{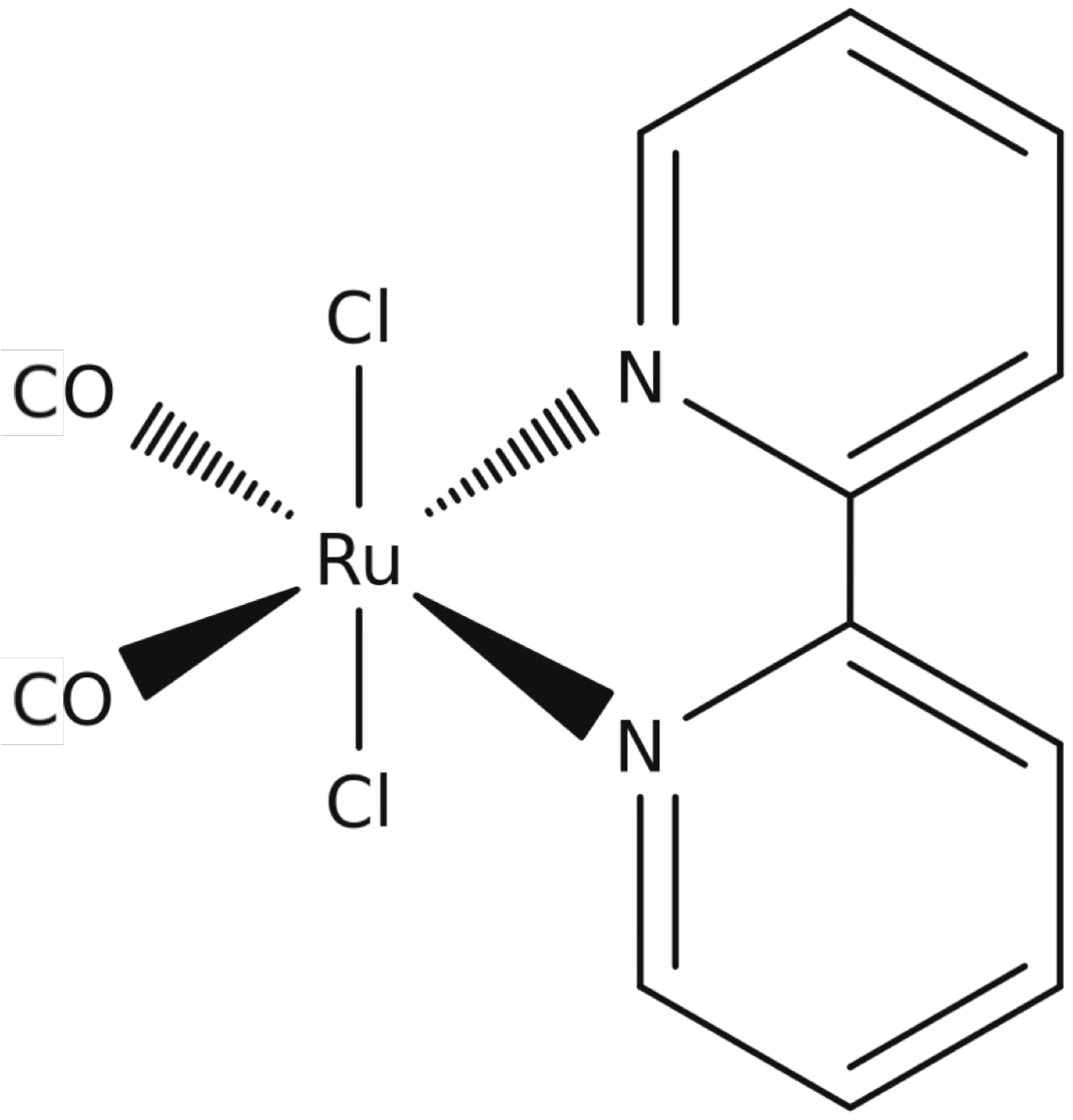}
    \caption{}\label{sfig:transcl_vsepr}
    \end{subfigure}
    \begin{subfigure}[b]{.35\linewidth}
\includegraphics[scale=0.27]{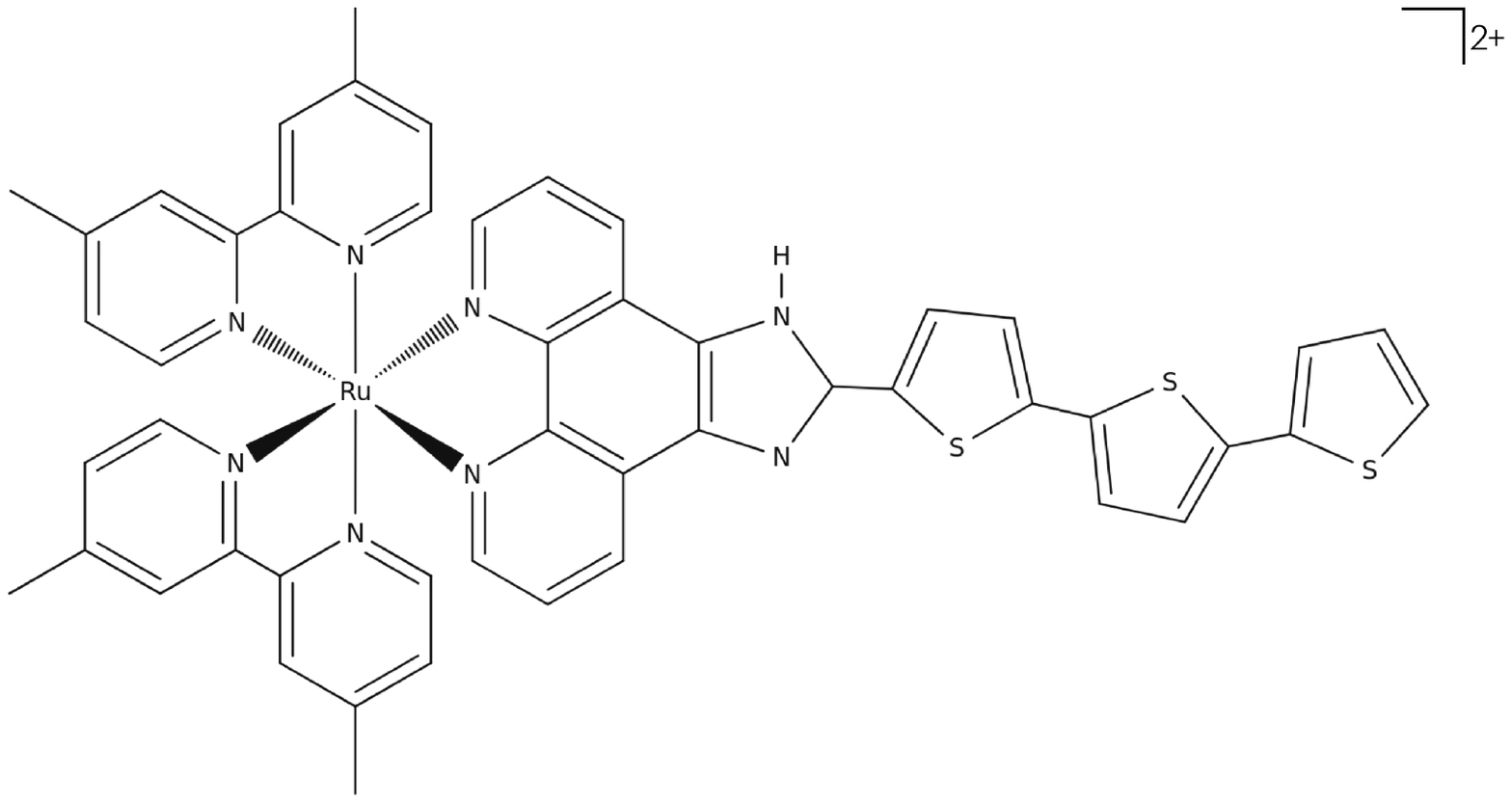}
    \caption{}\label{sfig:tld1433_vsepr}
    \end{subfigure}
    \caption{(\textbf{a}) Chemical structure of Trans-Cl (Trans(Cl)-Ru(bpy)Cl$_2$(CO)$_2$). (\textbf{b}) Chemical structure of TLD-1433 (rac-[Ru(dmb)$_2$(IP-3T)]Cl$_2$)~\cite{tld1433_fig}.}
\label{fig:tld1433_mol}
\end{figure}

Ruthenium(II) complexes constitute a prominent class of PDT molecules due to their favorable photophysical properties, including strong visible-light absorption, long-lived triplet excited states, and efficient generation of reactive oxygen species. Their excited-state manifolds often involve a complex interplay of metal-to-ligand charge-transfer (MLCT), ligand-centered, and metal-centered excitations, making them challenging targets for multi-configuration electronic-structure methods. As a result, the definition of balanced and chemically meaningful active spaces is essential for accurately describing their photophysics and photochemical reactivity, providing a stringent application benchmark for AEGISS. In case study, we have selected \teal{two} Ruthenium-based complexes models used for PDT, including one currently in clinical trials: Trans(Cl)-Ru(bpy)Cl$_2$(CO)$_2$ (dubbed in this work as \textit{Trans-Cl}, 27 atoms)
, and rac-[Ru(dmb)$_2$(IP-3T)]Cl$_2$ (\textit{TLD-1433}, 99 atoms), all displayed in Figure \ref{fig:tld1433_mol}.

The first serve\teal{s} as simplified prototype of photoactive drugs for PDT, with some literature available to guide our benchmarking of our semi-automated active-space selection approach. Trans-Cl, the smallest system, contains a single Ru(II)-coordinated ligand, making it useful for identifying key metal-to-ligand (M $\rightarrow$ L) transitions relevant to PDT. \teal{The latter,} TLD-1433\teal{,} presents the most challenging case due to its size and electronic structure complexity. With limited theoretical studies available, accurate wavefunction-based calculations remain unfeasible, leaving Time-Dependent DFT (TDDFT)~\cite{Casida2008, CASIDA2009, Runge1984} as our primary reference \cite{McFarland2019, Alberto2016}. 

\subsubsection{Trans-Cl}
\label{ssec:transcl_res}
\teal{This complex has been used as a playground to gain a better understanding of how to model Ru(II) complexes. Already in these prototypes, we can see that transitions of very different characters, i.e., intra-ligand (IL), metal-centered (MC), metal-to-ligand charge-transfer (MLCT), and ligand-to-metal charge-transfer (LMCT) states, can be found. Achieving a balanced description of these within a single active space is therefore non-trivial.}
\teal{To guide the AO-projection procedure, the molecule was partitioned into chemically meaningful regions. Four atom clusters were identified: $\mathcal{C}_0$, corresponding to the Ru(II) center; $\mathcal{C}_1$, corresponding to the bipyridine ligand; $\mathcal{C}_2$, containing the two CO groups; and $\mathcal{C}_3$, comprising the two chlorine atoms. Since the electronic transitions of interest predominantly involve the metal center and the bipyridine ligand, the orbital selection focused on clusters $\mathcal{C}_0$ and $\mathcal{C}_1$. The resulting partitioning is shown in Figure~\ref{fig:transcl_frags}.}

\begin{table}[!htb]
    \centering
    \begin{tabular}{c|c|c|c|c|c|c|c|c}
    \toprule
         Guess & Basis & CAS$_{\text{DMRG}}$ &BD& $\tau_E$ & $D$ & $n_D$ & CAS$_{A}$ & Name\\\hline\hline
         \multirow{3}{*}{\makecell{Canonical\\RHF}} & \multirow{3}{*}{\makecell{def-tZVP\\(cc-pVDZ-pp)$^*$}}   & \multirow{3}{*}{(92e,90o)} & \multirow{3}{*}{500} & \multirow{3}{*}{0.1} & \multirow{3}{*}{\makecell{$\mathcal{C}_0$:[Ru 4d$\textbf{x}$]\\$\mathcal{C}_1$:[C/N 2p$_z$]}} & \makecell{[1]\\(4,4)} & \multirow{3}{*}{(14e,13o)} & I \\\cline{7-7}\cline{9-9}
         & & & & & & \makecell{[2]\\ (6,6)} & & II \\
         \bottomrule
    \end{tabular}
    \caption{\teal{AEGISS settings for generating the Trans-Cl active-space. The initial guess orbitals are Canonical RHF with def-tZVP basis set for all the atoms and (*) Stuttgart-Dresden pseudo-potential on the metal. CAS$_{\text{DMRG}}$ corresponds to the $N_{\mathbf{A}}$ = 90, selected around the Fermi-level, while $\tau_E$ refers to 10\% of the $S(1)_{max}$, i.e. the maximum SOE amongst all the orbitals in the active subset $\mathbf{A}$. $n_D$ is the number of selected orbitals with highest $w_D$ contribution. With $[n]$, we refer to a selection in the whole subspace, while ($n_o$, $n_v$) is used to select the top $n_o$ and $n_v$, in the occupied and virtual space, respectively.}}
    \label{tab:sel_transcl}
\end{table} 
\teal{The AEGISS workflow described in Section~\ref{ssec:workflow} was applied using the computational setup reported in Section~\ref{apx:setup_transcl} and the selection parameters summarized in Table~\ref{tab:sel_transcl}.}

\teal{Space I was obtained by direct application of the AEGISS. The active space contains the five Ru(II) $\mathrm{4d}$ orbitals, selected by retaining the dominant molecular orbital contribution associated with each Ru $\mathrm{4d}$ component. This was enforced through the criterion $n_{[\mathrm{Ru};4d\mathbf{x}]}=[1]$, corresponding to the highest-ranked orbital for each Ru $\mathrm{4d}$ contribution. The selected set comprises three occupied orbitals, predominantly of $\mathrm{d}_{xy}$, $\mathrm{d}_{xz}$, and $\mathrm{d}_{yz}$ character, and two virtual orbitals of mainly $\mathrm{d}_{z^2}$ and $\mathrm{d}_{x^2-y^2}$ character. To complement the metal-centered orbitals, a balanced set of bipyridine $\pi/\pi^\ast$-orbitals was selected using the AO label $[\mathrm{C/N}\;\mathrm{2p}_z]$. Occupied and virtual subspaces were treated separately, retaining the four highest-ranked orbitals in each subset. 
The resulting Space I defines a CAS(14e,13o) active space generated directly from the AEGISS pipeline. The automatically generated candidate space already captures the chemically relevant orbital manifold.}

\begin{figure}[!hbt]
    \centering
\includegraphics[width=0.7\linewidth,trim={0 3.5cm 0 2.7cm}, clip]{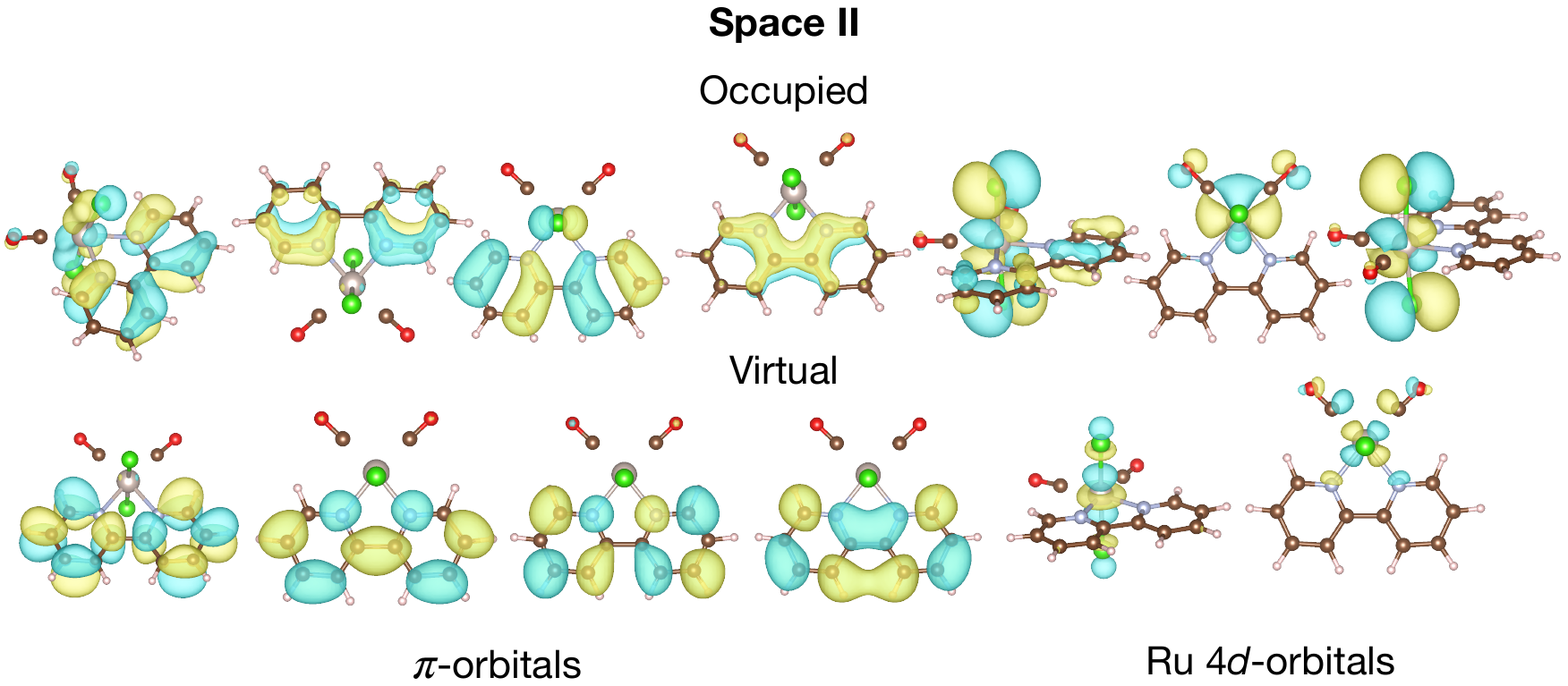}
    \caption{\teal{Final active space identified by AEGISS procedure for the Trans-Cl molecule, corresponding to space II. From the selected orbitals is possible to group them according to $\pi$-character, like the first four of each row as well as the remaining five orbitals, corresponding to the Ruthenium 4ds.}}
    \label{fig:TransCl_spaceII}
\end{figure}

\teal{Space II was constructed using a less restrictive selection criterion. In this case, the two highest-ranked orbitals associated with each Ru $4d$ component were retained, together with six occupied and six virtual orbitals selected from the $[\mathrm{C/N};2p_z]$ label. Under these conditions, an ambiguity arises regarding the inclusion of the HOMO and HOMO$-1$ orbitals. Although these orbitals exhibit significant Ru $\mathrm{4d}_{xz}$ and $\mathrm{4d}_{yz}$ character, comparable to that of HOMO$-13$ and HOMO$-12$, they possess slightly lower single-orbital entropy values. Space II was therefore generated from Space I by replacing HOMO$-13$ and HOMO$-12$ with HOMO$-1$ and HOMO, corresponding to the second-largest contributions associated with the Ru $\mathrm{4d}_{xz}$ and $\mathrm{4d}_{yz}$ labels. Following the refinement procedure, the orbital set was reduced to the same CAS(14e,13o) size as Space I. Space~II explicitly includes the HOMO and HOMO$-1$ orbitals and yields excitation energies, orbital shapes and symmetries, and state characters in agreement with the literature (Table~\ref{tab:transcl_casscf} and Figure~8 in the SI). The resulting active space required only limited refinement before the final multi-reference calculations. For both Space I and Space II, the distribution of irreducible representations matches that reported in the literature, Ref.~\cite{TransCl}, namely two A$_1$, one B$_1$, five A$_2$, and one B$_2$ orbitals.}

\teal{We applied the AVAS algorithm using the same AO labels reported in Table~\ref{tab:sel_transcl}. By construction, AVAS does not allow distinct AO label sets to be assigned to different regions of the molecule: a single projector is built from the union of the requested labels. As a consequence, AVAS attempts to define orbitals that project simultaneously onto the Ru 4d shell and onto the bpy $\pi/\pi^*$ network. The resulting active spaces, CAS(22,17) at cutoff 0.2 and CAS(20,15) at cutoff 0.5, contain orbitals of strongly mixed metal–ligand character. This hybridization is an artifact of the single-projector construction rather than a physical feature of the electronic structure. In Trans-Cl the Ru-centered and bpy-centered correlation channels are more naturally described as separable sets of orbitals, which is the regime AEGISS is designed to handle by selecting orbitals on a per-region basis.}

\subsubsection{Spin-Averaged CASSCF benchmark}
\teal{Previous CASSCF studies of Trans-Cl have shown that the low-lying singlet manifold is dominated by metal-centered (MC) states, corresponding to $S_1$-$S_5$ and $S_7$~\cite{TransCl}. In contrast, the lowest singlet states with significant oscillator strengths are the MLCT and IL states $S_8$, $S_9$, $S_{10}$, and $S_{11}$. The MLCT states primarily involve excitations from Ru $4d$ orbitals to bipyridine $\pi^\ast$-orbitals, whereas the IL states arise from transitions within the ligand $\pi/\pi^\ast$-manifold. A balanced description of these excitations therefore requires the inclusion of all five Ru $4d$ orbitals together with a suitable set of ligand $\pi/\pi^\ast$ orbitals and a sufficiently large number of excited-state roots. Experimentally, the UV-vis spectrum measured in CH$_3$CN exhibits a weak absorption band centered at 352 nm (3.52 eV), corresponding to 3.62 eV in CH$_2$Cl$_2$. In Ref.~\cite{TransCl}, this transition was assigned to the $S_9$ state at the CASSCF level but becomes the lowest singlet excited state after inclusion of dynamical correlation through CASPT2. This reordering highlights the importance of dynamical correlation effects in determining the relative energetics of the excited-state manifold. The states identified as most relevant for the interpretation of the experimental spectra are the MLCT states 4$^{1}$B$_2$ and 3$^{1}$A$_1$, the IL state 5$^{1}$B$_2$, and the MC state 1$^{1}$B$_1$.}

\teal{To assess the quality of the generated active spaces, spin-averaged (SA)-CASSCF calculations were performed and compared with the reference CASSCF results. More sophisticated treatments, such as CASPT2, were not considered in the present work. }

\teal{Using the same number of singlet (12) and triplet (11) roots employed in Ref.~\cite{TransCl}, both Space I and Space II reproduce the main qualitative features of the low-lying excited-state manifold. In particular, the lowest excited states exhibit predominantly metal-centered character, while the state corresponding to $S_9$ displays metal-to-ligand charge-transfer character, consistent with the assignments reported in the reference study. The two active spaces differ primarily in the treatment of the Ru-centered orbitals associated with the HOMO/HOMO$-1$ pair. Space II explicitly includes these orbitals and yields excitation energies, shape and symmetries of the orbitals in the active space and state characters which are in agreement with the literature, see Figure~\ref{fig:TransCl_spaceII}.} 
Table~\ref{tab:transcl_casscf} reports the SA-CASSCF excitation energies obtained from space~II alongside the reference values of Ref.~\cite{TransCl}, as well as the TD-DFT results from two functionals, PBE0 and B3LYP.

\teal{Notably, the TD-DFT excitation energies are in excellent agreement with the corresponding reference values. A more extensive investigation with both CASSCF and TDDFT is provided in the Supporting Information (Section C) in Table~\ref{tab:transcl_casscf_full}, for CASSCF, while  
Table~\ref{tab:transcl_tddft_functionals_s} and Table~\ref{tab:transcl_tddft_functionals_t}, for TD-DFT . The remaining quantitative differences observed for the multi-configurational calculations are attributable to methodological differences between the two studies, including the use of different basis sets, software packages, and computational protocols. In particular, the reference work employed the ANO-RCC-VTZP basis set together with Gaussian09 and MOLCAS 7.2, whereas the present calculations were performed with PySCF and the closely related def2-TZVP basis set.}

\begin{table}[!htb]
\centering
\setlength\tabcolsep{4pt}
\begin{tabular}{c|c|c|cc}
\toprule
 State& SA-CASSCF (II) & SA-CASSCF~\cite{TransCl}& PBE0 & B3LYP\\\hline
1$^1$B$_1$ & 3.51 (S$_2$) &3.59  & 3.30 & 3.25 \\ 
3$^1$A$_1$& 6.60 (S$_9$) & 5.61& 2.58 & 2.40 \\ \bottomrule
\end{tabular}
\caption{\teal{Two singlets of interest, 1$^1$B$_1$ and 3$^1$A$_1$, excited state transition energies (in eV) relative to $\text{S}_0$, obtained from SA-CASSCF (12 singlets and 11 triplets) calculations on the Trans-Cl complex with active space II.  }}
\label{tab:transcl_casscf}
\end{table}

\subsubsection{TLD-1433}
\teal{For this molecule, which exhibits a well-separated structural organization, we adopted an atomic-labeling strategy similar to that used for the trans-Cl Ru(II) complex. The system was partitioned into three principal clusters: $\mathcal{C}_0$, containing the Ru(II) center; $\mathcal{C}_1$, comprising the bipyridine ligand extended by the thiophene chain; and $\mathcal{C}_2$, formed by the two remaining bipyridine ligands. This fragmentation reflects the dominant chemical motifs involved in the low-lying electronic excitations. Nevertheless, a more refined partitioning could readily be adopted by further decomposing the elongated ligand, for example by assigning individual clusters to each thiophene ring and an additional cluster to the imidazole moiety that connects the ligand tail to the metal center.}

\begin{figure}[!htb]
    \centering
    \includegraphics[width=0.65\linewidth]{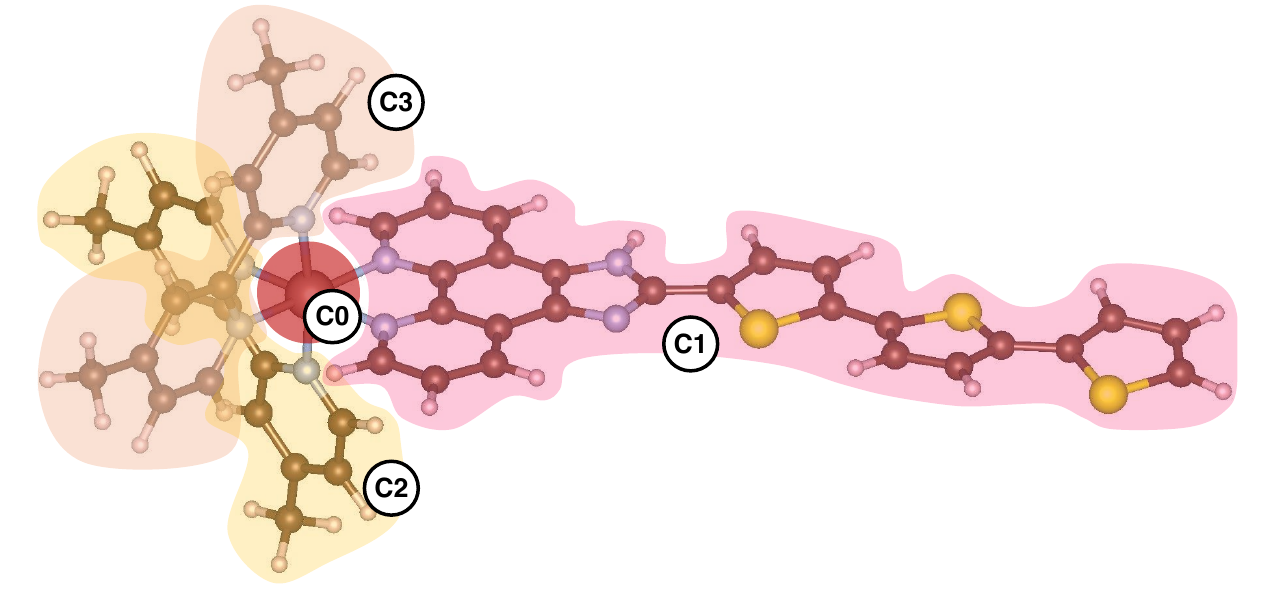}
    \caption{TLD-1433 molecule with highlighted cluster defined according to the coarse atom-labeling, as explained in this section. Each of the ligand (including the elongated one) is treated as an independent cluster, as the Ru(II) atom.}
    \label{fig:tld1433_frags}
\end{figure}

\teal{To the best of our knowledge, no active-space definitions have been reported in the literature for this system within CASCI- or CASSCF-based frameworks. We therefore begin from an active-space construction analogous to that employed for the trans-Cl Ru(II) complex and use AEGISS to systematically extend the orbital manifold toward larger and increasingly correlated active spaces. This approach enables the exploration of active spaces that would be impractical for standard CASCI/CASSCF treatments, while preserving a chemically motivated description of the relevant electronic states. The resulting hierarchy of active spaces is intended not only for classical electronic-structure calculations but also for the generation of effective Hamiltonians suitable for quantum computing applications~\cite{ADAPT_og,ADAPTVMPE}.}

\teal{We applied the AEGISS workflow} as outlined in Section \ref{ssec:workflow}, \teal{using the selection setting of Table~\ref{tab:sel_tld}, starting from the computational setup reported in Section~\ref{apx:tld1433_setup}.}
\begin{table}[!htb]
    \centering
    \begin{tabular}{c|c|c|c|c|c|c|c|c}
    \toprule
         Guess & Basis & CAS$_{\text{DMRG}}$ &BD& $\tau_E$ & $D$ & $n_D$ & CAS$_{A}$ & Name\\\hline\hline
         \multirow{5}{*}{\makecell{Canonical\\RHF\\---\\H$_2$O\\C-PCM}} & \multirow{5}{*}{\makecell{def-sVDP\\(cc-pVDZ-pp)$^*$}}   & \multirow{5}{*}{(80e,100o)} & \multirow{5}{*}{500} & \multirow{5}{*}{0.1} & \multirow{5}{*}{\makecell{$\mathcal{C}_0$:\texttt{"Ru 4d"}\\$\mathcal{C}_1$:\texttt{["C/N 2pz"]}\\$\mathcal{C}_2$:\texttt{["C/N 2p"]}\\$\mathcal{C}_3$:\texttt{["C/N 2p"]}}} & \makecell{(5,5)\\(1,1)\\(1,1)\\(1,1)} & \multirow{5}{*}{(14e,13o)} & I\\\cline{7-7}\cline{9-9}
         & & & & & & \makecell{(7,7)\\ (2,2)\\ (2,2)\\ (2,2)} & & II\\
         \bottomrule
    \end{tabular}
    \caption{\teal{AEGISS settings for the generating the TLD-1433 active-space. The initial guess orbitals are Canonical RHF with def-tZVP basis set for all the atoms and (*) Stuttgart-Dresden pseudo-potential on the metal, and C-PCM~\cite{TRUONG1995253, Barone1998} for taking into account WATER solvation effects. CAS$_{\text{DMRG}}$ corresponds to the $N_{\mathbf{A}}$ = 80, selected around the Fermi-level, while $\tau_E$ refers to 10\% of the $S(1)_{max}$, i.e. the maximum SOE amongst all the orbitals in the active subset $\mathbf{A}$. $n_D$ is the number of selected orbitals with highest $w_D$ contribution. With $[n]$, we refer to a selection in the whole subspace, while ($n_o$, $n_v$) is used to select the top $n_o$ and $n_v$, in the occupied and virtual space, respectively.}}
    \label{tab:sel_tld}
\end{table} 

\teal{A balanced active space for describing inter-ligand (IL) and metal-to-ligand charge-transfer (MLCT) states should include the Ru(II) 4d orbitals ($\mathcal{C}_0$), a balanced set of $\pi/\pi^*$-orbitals on the thiophene tail ($\mathcal{C}_1$), the $\pi$ orbitals on the two non-substituted bipyridine ligands coordinating the metal center ($\mathcal{C}_2$ and $\mathcal{C}_3$), and, when needed, $\sigma$-orbitals describing the bond between the thiophene tail and the coordinating bipyridine ligand. Exploratory calculations, similarly to DOBNA, revealed Rydberg-like orbitals among the first 20 virtual orbitals, together with strong mixing between Ru 4d and bipyridine character. To account for this metal-ligand hybridization, we therefore adopted a conservative selection strategy based on a fixed number of orbitals per AO label, rather than a threshold-based criterion.}

Following these considerations, a hierarchy of AO-projection schemes can be defined. The smallest chemically motivated active space corresponds to a \textbf{CAS(16,16)}, comprising one $\pi/\pi^*$ pair from each of the $\mathcal{C}_2$ and $\mathcal{C}_3$ ligands, five occupied and five virtual Ru 4d orbitals, and one $\pi/\pi^*$ pair on the thiophene tail. As in the previous systems, the goal is to obtain a balanced description of the low-lying states through a representative set of Ru-centered and ligand-centered orbitals. Inspection of the resulting molecular orbitals revealed that the final active space can be further reduced for two reasons:
\begin{enumerate}
    \item Due to the similarity of the two non-substituted bipyridine ligands, the same molecular orbital selected through AO projection often provides the dominant $\pi$ or $\pi^*$ contribution for both $\mathcal{C}_2$ and $\mathcal{C}_3$.
    \item The delocalized nature of the RHF orbitals leads to significant mixing between Ru 4d and ligand character, particularly in the virtual space, such that a single molecular orbital may simultaneously capture contributions from multiple AO labels.
\end{enumerate}

\begin{figure}[!hbt]
    \centering
    \includegraphics[width=0.8\linewidth,trim={0 2.5cm 0 2.7cm}, clip]{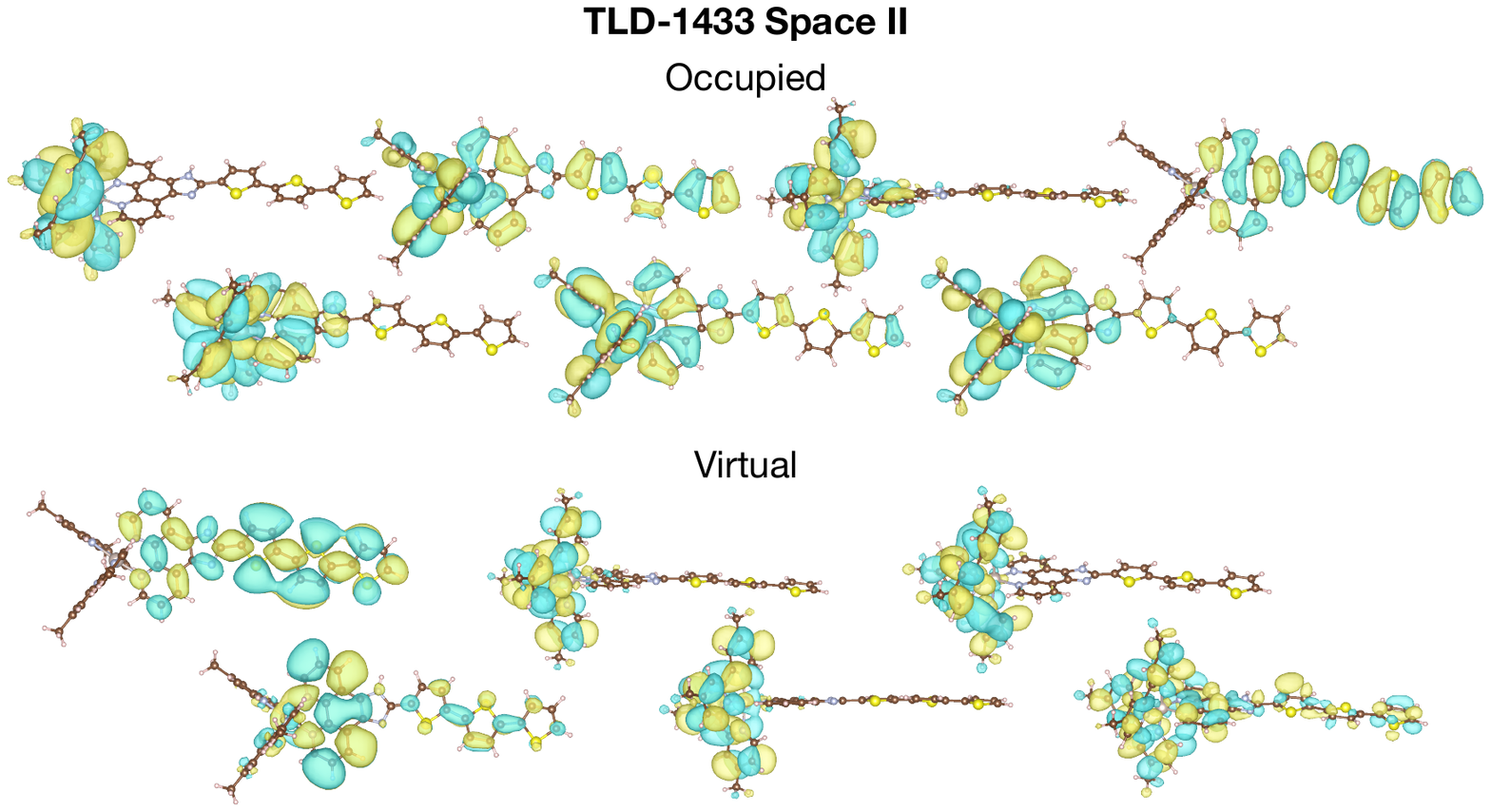}
    \caption{\teal{Final active space identified by AEGISS procedure for the TLD molecule, corresponding to space II. From the selected orbitals is possible to group them according to $\pi$-character, like the first four of each row as well as the remaining five orbitals, corresponding to the Ruthenium 4ds.}}
    \label{fig:TLD_spaceII}
\end{figure}

As a result, the nominal \textbf{CAS(16,16)} naturally reduces to a \textbf{CAS(14,13)}. However, this out-of-the-box active space, space I, includes a Rydberg-like virtual orbital and excludes the HOMO. Increasing the selection to two orbitals per $\pi/\pi^*$ AO label and seven Ru 4d orbitals provides a refined candidate space, from which the Rydberg orbital can be removed and the HOMO restored, yielding a more balanced description of the MLCT and IL states. \teal{The second space (II) is shown in Figure~\ref{fig:TLD_spaceII}.}

\teal{Given the observations made on Trans-Cl, and the structural analogy of TLD-1433, i.e. a Ru center coupled to an even more extended ligand $\pi/\pi^*$-manifold, we expect the single-projector limitation to manifest at least as severely on this system.}

For TLD-1433, we focused on Space II, which was treated with PCM-DMRG-CI, as PCM-CASSCF calculations become prohibitively expensive for a system of this size. \teal{The details of the calculation are reported in Section~\ref{apx:tld1433_setup}.} \teal{The PCM-DMRG-CI total energies for $S_0$, $S_1$ and $T_1$, together with the corresponding $S_0 \to S_1$ and $S_0 \to T_1$ vertical gaps for Space~II, are reported in Table~\ref{tab:TLD1433_gaps}}. 

\begin{table}[!htb]
    \centering
    \begin{tabular}{c|c|c|c|c|c|c}
    \toprule
         \textbf{AS} & \textbf{S$_0$} & \textbf{S$_1$} & \textbf{$\Delta($S$_1 - $S$_0)$} & \textbf{T$_1$} & \textbf{$\Delta($T$_1 - $S$_0)$}& \textbf{$\Delta($S$_1 - $T$_1)$} \\ \hline\hline
         II & -3597.6059& -3597.432& \makecell{4.719 \\ (0.173)} & -3597.461 & \makecell{3.935
\\(0.145)} & \makecell{
0.785\\(0.029)}\\ \bottomrule
    \end{tabular}
    \caption{S$_0$, S$_1$, and T$_1$ obtained from a PCM-DMRGCI calculation with space II defining a CAS(14,13). Energies are given in Hartree. The gaps are given in eV and in brackets are reported the conversion in Hartree.}
    \label{tab:TLD1433_gaps}
\end{table}

Given the absence of established computational benchmarks for this molecule based on wave function methods, we are also conducting a thorough DFT investigation, the results of which are provided in the SI. This detailed analysis offers a reliable starting point and a potential reference for future high-accuracy calculations. 

\section{Discussion and future perspectives}
\label{sec:conclusion}
\teal{The results presented throughout this work show that AEGISS provides a practical workflow for constructing active spaces across a diverse range of chemically and electronically challenging systems. From benchmark molecules such as benzene and ferrocene to a representative TADF emitter and increasingly complex Ru(II)-based photosensitizers, the workflow consistently identifies active spaces that support accurate multireference calculations. Rather than eliminating user intervention entirely, AEGISS aims to reduce reliance on trial-and-error procedures by providing a systematic starting point that can be refined when required for particularly challenging electronic structures.}

\teal{The central idea behind AEGISS is that active-space selection requires answering two complementary questions: where strong correlation resides and which of those correlated orbitals are chemically relevant to the process of interest. AEGISS addresses these questions through single-orbital entropies and atomic-orbital projections, respectively, integrating complementary ideas from AutoCAS and AVAS into a unified workflow. Neither descriptor alone fully addresses the active-space selection problem: entropy identifies correlated orbitals but does not distinguish those most relevant to the targeted chemistry, whereas AO projections preserve chemical meaning but cannot identify strong correlation by themselves. Their combination therefore enables the construction of compact, chemically motivated active spaces. A further advantage of the workflow is that  distinct AO groups can be treated independently, allowing different correlation channels—for example metal-centered and ligand-centered orbitals—to be selected separately rather than through a single global projection. The Ru(II) case studies illustrate the practical benefit of this strategy, where independent treatment of metal and ligand orbitals facilitates the construction of balanced active spaces while preserving chemical interpretability. At the same time, AEGISS intentionally adopts a conservative selection strategy that minimizes the risk of excluding chemically important orbitals. Expert inspection therefore remains beneficial for systems exhibiting extensive orbital mixing, Rydberg contamination, or multiple chemically equivalent orbital choices. The modular separation between the computationally demanding entropy analysis and the comparatively inexpensive projection stage further means that incorporating chemically guided orbital selection adds negligible computational cost once the correlated calculation has been performed.}

\teal{Beyond classical multireference calculations, AEGISS was developed with quantum computing applications in mind, where active-space construction represents a critical preprocessing step. By generating compact, chemically motivated active spaces, the workflow facilitates the construction of molecular Hamiltonians suitable for both near-term variational algorithms, such as the Variational Quantum Eigensolver (VQE)~\cite{vqe_rev}, and fault-tolerant approaches including Quantum Phase Estimation (QPE)~\cite{kitaev1995qpe,nielsen2010}. This capability has already been demonstrated in our recent ADAPT-VMPE study~\cite{ADAPTVMPE} on a close derivative of TLD-1433, where active spaces generated with AEGISS were used to construct excited-state Hamiltonians for deployment on the Aurora quantum computing platform~\cite{Q4Bio}.}

\teal{Several directions remain for future development, including further automation of orbital identification, the exploration of alternative correlation metrics~\cite{Ding_2023,evangelista2025}, machine-learning-assisted pre-screening strategies~\cite{golub2020}, and synergies with emerging approaches such as the Active-Space Finder (ASF)~\cite{asf,zhao2025}. More generally, AEGISS should be viewed as a flexible workflow that can naturally incorporate future advances in correlated electronic-structure methods and automated orbital selection. We anticipate that workflows of this kind will facilitate both classical multireference calculations and the preparation of compact molecular Hamiltonians for quantum computing applications.}

\subsection*{Acknowledgements}
Work on “Quantum Computing for Photon-Drug Interactions in Cancer Prevention and Treatment” is supported by Wellcome Leap as part of the Q4Bio Program. 
This work made use of the High Performance Computing Resource in the Core Facility for Advanced Research Computing at Case Western Reserve University and LEONARDO supercomputer, owned by the EuroHPC Joint Undertaking, hosted by CINECA (Italy). LG and FT acknowledge funding from the European Union - Next Generation EU, Mission 4 - Component 1 - Investment 4.1 (CUP E11I22000150001). LG acknowledges funding from the Ministero dell’Universit\'a e della Ricerca (MUR) under the Project PRIN 2022 number 2022W9W423. VK acknowledges funding support from Cleveland Clinic Research.

\subsubsection*{Supporting Information Available}
The supporting information includes: optimized geometries for the molecules investigated, classical benchmarks calculations, in particular, TDDFT and SA-CASSCF when available. Outputs and inputs for the calculations presented in this work are deposited on Github (https://github.com/MartStella/AEGISS/) and can be obtained upon reasonable request.

\subsubsection*{Conflict of interest}
The authors declare no conflict of interest.

\subsubsection*{Author contributions}
Author contributions are listed according the CRediT (Contributor Roles Taxonomy) classification.\\
\textit{Funding Acquisition:} SK, VK, LG. 
\textit{Conceptualization:} PABH, FT, MS.
\textit{Data Curation:} FT, MS.
\textit{Methodology:} PABH, FT, MS.
\textit{Software:} PABH, FT.
\textit{Investigation:} FT, MS, PABH, SK, FP, VK. 
\textit{Supervision:} SK, FP, MS, LG.
\textit{Visualization:} FT, MS.
\textit{Writing – original draft:} FT, MS. 
\textit{Writing – review \& editing:} All authors.
\nocite{avogadro, vesta2008, psiemb, zhai2023block2, B3LYP_1, B3LYP_2, def2tzvp, pp_transcl, ORCA, ano_rcc_tzvp_1, ano_rcc_tvzp_2, RI-MP2, g16, PBE0_1, PBE0_2, TRUONG1995253, Barone1998, def2svpd, bogu15a}

\printbibliography
\clearpage
\appendix
\input{SI}
\end{document}

%% file: SI.tex
\title{Supplementary Information\\
AEGISS - Atomic orbital and Entropy-based Guided Inference for Space Selection - 
A novel semi-automated active space selection workflow for quantum chemistry and quantum computing applications\\\textbf{Supplementary Information}}

\maketitle
\section{How to plug a Quantum Chemistry problem into a Quantum Computer}

We begin with the electronic Hamiltonian in second quantization, expressed over the full molecular orbital (MO) basis:

\begin{equation}
\hat{H} = \sum_{pq} h_{pq} \, a_p^\dagger a_q + \frac{1}{2} \sum_{pqrs} (pq|rs) \, a_p^\dagger a_q^\dagger a_s a_r
\end{equation}

\noindent where $h_{pq}$ are the one-electron integrals,  $(pq|rs)$ the two-electron integrals (in chemist's notation),  $a_p^\dagger, a_q $ fermionic creation and annihilation operators.

The active space selection procedure allows to define a subset of molecular orbital, $\mathcal{A} \subset \mathcal{M}$, from the total molecular orbital space, $ \mathcal{M}$. Thus, the MO basis can be partitioned into \textit{core orbitals}, $\mathcal{C}$, orbitals that are going to be always doubly occupied, the \textit{active orbitals}, $\mathcal{A}$, which are going to be treated more accurately either classically or by a quantum computing method, and the \textit{virtual orbital}, $\mathcal{V}$, which are going to be always unoccupied.
The core orbitals indexed by $u, v \in \mathcal{C} $, active orbitals indexed by $i, j, k, l \in \mathcal{A} $, and virtual orbitals indexed by $a, b, c, d \in \mathcal{V} $.

Following this subdivision, the effective one-body and two-body integrals can be defined, namely $h_{pq}^{eff}$ and $\Gamma_{pqrs}^{eff}$, as well as the new core energy, $E_{core}$. Restricting to the active space, the\textit{effective} Hamiltonian, $H_{eff}$, is defined as
\begin{equation}
\hat{H}_{eff} = E_{\text{core}} + \sum_{ij \in \mathcal{A}} h^{\text{eff}}_{ij} \, a_i^\dagger a_j + \frac{1}{2} \sum_{ijkl \in \mathcal{A}} \Gamma_{ijkl}^{eff} \, a_i^\dagger a_j^\dagger a_l a_k.
\label{eq:H_elec}
\end{equation} 
In order to obtain the new effective component, the following contraction are required. Assuming the core orbitals are always doubly occupied, their contribution to the energy becomes a constant scalar term
\begin{equation}
E_{\text{core}} = \sum_{u \in \mathcal{C}} h_{uu} + \frac{1}{2} \sum_{u, v \in \mathcal{C}} \left[ 2(uu|vv) - (uv|vu) \right]. 
\end{equation}The one-electron integrals over the active space are modified due to the presence of the frozen core, leading to the effective one-body integrals as
\begin{equation}
h^{\text{eff}}_{ij} = h_{ij} + \sum_{u \in \mathcal{C}} \left[ 2(ij|uu) - (iu|uj) \right], \quad i, j \in \mathcal{A},
\end{equation} so by reducing two-body interaction to one-body interaction restricted to the active space. The effective-two body tensors instead is just obtained as $\Gamma^{eff}_{ijkl} = (ij|kl)$, where the indices $i,j,k,l \in \mathcal{A}$, leaving outside all the indices non included in $\mathcal{A}$.
Without building the effective Hamiltonian, the effective integrals can be now be used by a standard \textit{classical} (used improperly) quantum chemistry pipeline in order to apply one of the methods listed in the main manuscript (e.g. CASCI, CASSCF, DMRG, CASPT2).

If the final goal instead is to perform a calculation on a quantum computer, so by exploiting a quantum algorithm, the first essential step is to performs a mapping between the Fermionic space to the qubit space by means of a Fermionic-to-Qubit mapping method. Different method exist to convert Fermionic operators to the bosonic space of the qubits, above all the most popular are Jordan-Wigner~\cite{jordan1928pauli}, Parity~\cite{parity_2017}, Bravy-Kitaev~\cite{bk_2002}, but also custom and ad-hoc mappings can be defined ~\cite{Miller2023, Miller2024}.
Here, we proceed the explanation by exploiting the Jordan-Wigner mapping. The Jordan-Wigner mapping allows for a more intuitive transformation, maintaining a almost 1-to-1 correspondence between Slater determinants, and states in the computational basis, so in the qubit space. Each active spin orbital is going to be mapped to one qubit and the encoding represent the occupation of that specific spin orbital, as for Slater Determinants. The ordering of Fermionic modes and qubits is not fixed, changing from software to software and from publication to publication, but generally the two main orderings are \textit{splitted}, $\ket{0_{\beta}\dots0_\beta0_\alpha\dots0_\alpha}$, thus splitting alpha and beta spin orbitals, or \textit{interleaved}, $\ket{0_\beta0_\alpha\dots0_\beta0_\alpha}$. For this explanation, we stick to the first one. The \textit{vacuum} state is going to be defined in the same way, i.e. $\ket{\mathbf{0}}_f = \ket{0\dots00}_f \rightarrow \ket{\mathbf{0}}_q = \ket{0\dots00}_q  $. The Hartree-Fock state of an H$_2$ two orbital system, $\ket{HF} = \ket{0101}_f$, is going to be mapped to the computational basis vector $\ket{HF}_q = \ket{0101}_q$. An excited determinant is going to follow the same structure, e.g. $\ket{1001}_f \rightarrow \ket{1001}_q$. 
The transformation applies to the Fermionic operators. In particular, 
\begin{equation}
\hat{a}^{\dagger}_{i} = \hat{Q}_i^{\dagger}\prod^{i-1}_{j=0}{Z_j} \quad \text{and} \quad
\hat{a}^{}_{i} = \hat{Q}^{}_i\prod^{i-1}_{j=0}{Z_j},
\label{eqn:aa+}\end{equation} where $\hat{Q}^{\dagger}_{i} = \frac{1}{2}(X_i - iY_i)$, $\hat{Q}_{i} = \frac{1}{2}(X_i + iY_i)$, and $i$ iterates over the set of active spin orbitals. The $\hat{Q}$ operators can be associated to creation and annihilation operators in the qubit space, respectively. As well as the Fermionic creation and annihilation operators act to modify the occupancy of the relative spin-MO, the qubit operators act to change the state of the qubit related to the specific spin orbital. The full correspondence is finalized with a series of Pauli-Z matrices that allow to compute the parity of the state and account for the Fermionic anticommutation between $a$ and $a^{\dagger}$. Using Equations \ref{eqn:aa+} in Equation \ref{eq:H_elec}, the effective Hamiltonian in the qubit space, $\hat{H}_q$, can be written as 
\begin{equation}
\hat{H_q} = \sum_i^L c_i \hat{P}_i = \sum_{i}^L c_i \bigotimes_{j=0}^{N_{MO}}{\hat{\sigma}_j^i},
    \label{eqn:H_elec_qubit}
\end{equation} where each $\hat{\sigma}_j \in \{X,Y,Z,I\}$ is a Pauli matrix acting on the $j$-th qubit, $c_i$ is a coefficient relative to the $i$-th Pauli string $\hat{P}_i$, $N_{MO}$ is the number of qubit or number of spin-MOs, and $L$, proportional to $N_{MO}^4$, is the number of Pauli string composing the qubit Hamiltonian. Each Pauli string $\hat{P}_i$ is obtained by the multiplication of each term in the original Hamiltonian, $\hat{H}_{eff}$, once they get mapped in the qubit space.

\section{\teal{Atomic-Labelling Guidelines}}
\label{apx:atom-labeling}
\teal{This section provides practical recipes for the atom-labeling step (step~\textcircled{1} of Section~3) for users approaching a new system. The clustering does not affect the entropy diagnostic of step~\textcircled{3}; it only controls how AO labels are distributed across the molecule in step~\textcircled{4}.
}
\teal{The molecule is partitioned into clusters $\{C_i\}$ following the \textit{spatial locality of the targeted electronic process}. Atoms that play distinct chemical roles in the process of interest - donor vs.\ acceptor, metal vs.\ ligand, photo-active centre vs.\ spectator - are placed in distinct clusters. Atoms whose role is chemically equivalent or known to be inactive can be grouped together, or omitted from the AO-label specification entirely.
}

\noindent\teal{\textbf{Localised processes
($d$-$d$, $\pi \to \pi^*$, bond dissociation).}}
\teal{A two- or three-cluster partitioning is usually sufficient: one cluster encompassing the active site (the transition metal, the aromatic ring, the dissociating bond), and one or two further clusters grouping the surrounding ligands or substituents. AO labels in step~\textcircled{4} are specified only on the active cluster. \textit{Example}: ferrocene (Section~4.2) uses a single cluster centred on Fe with AO label ``Fe $3d$''; the cyclopentadienyl rings are present in the candidate pool through the DMRG entropy step but receive no AO label of their own.
}

\noindent\teal{\textbf{Delocalised or charge-transfer processes
(MLCT, LMCT, donor-acceptor CT).}}
\teal{ Both the donor and the acceptor moieties are placed in distinct clusters, so that per-cluster AO labels target each side independently. \textit{Example}: trans-Cl (Section~4.4.1) uses clusters $C_0$ (Ru), $C_1$ (bipyridine), and $C_2$ (two CO ligands), with AO labels ``Ru $4d$'' on $C_0$ and ``C/N $2p_z$'' on $C_1$; this partitioning naturally produces a balanced metal-vs-ligand active space appropriate for MLCT and MC transitions.}

\noindent\teal{\textbf{Extended or polycyclic ligands.}}
\teal{ For ligands whose internal structure may itself influence the active space (multiple aromatic rings, conjugated linkers), a finer partitioning assigns one cluster per ring or per chemically distinct fragment. \textit{Example}: TLD-1433 (Section~4.4.3) uses four clusters ($C_0$ = Ru, $C_1$ = thiophene tail, $C_2$ and $C_3$ = the two non-substituted bipyridine ligands). A finer alternative, in which the thiophene tail is itself decomposed into individual thiophene rings and an imidazole linker, is also discussed in Section~4.4.3.}

\noindent\teal{\textbf{No prior chemical knowledge.}}
\teal{ If the user cannot anticipate where the relevant orbitals are localised, the clustering step can be skipped entirely. AEGISS will then perform AO projection on the molecule as a whole, with AO labels applied indiscriminately across all atoms of the same element. This is the AVAS-like limit of the workflow and is appropriate for small or symmetric systems where chemical intuition does not provide a useful partitioning.
}

\noindent\teal{\textbf{Coarse-to-fine refinement}}
\teal{ In ambiguous cases the recommended workflow is to start from a coarse partitioning (one cluster per chemically distinct sub-structure) and refine only if the resulting active space shows an imbalance - e.g.\ one fragment dominating the AO projection while another is under-represented. Refining the clustering is computationally inexpensive, since the DMRG entropy extraction in step~\textcircled{3} does not need to be re-run: only the AO projection in step~\textcircled{4} is repeated with the new label set on the new cluster definition.
}

\noindent\teal{\textbf{Impact of cluster choice}}
\teal{ The cluster boundaries determine, via the per-cluster AO labels, which orbitals are admitted by the AO-projection step. Coarser clusters force the same AO labels to apply uniformly across larger spatial regions, which can over-include orbitals on chemically inactive sub-units. Finer clusters allow targeted selection at the cost of additional user input. The clustering does \textit{not} affect the entropy-screened candidate pool, so mis-specifying the clustering does not exclude correlated orbitals from consideration; in the worst case, it produces a slightly less compact active space that can be tightened by re-running step~\textcircled{4} with refined labels.
}
\section{Single orbital entropy: Alternative definition and sources}
\label{apx:alt_soe}

\teal{The one-orbital RDM defined in Equation~\ref{eqn:1o-rdm} is diagonal in the local occupation basis $\{\ket{0},\ket{\uparrow} \ket{\downarrow},\ket{2}\}$ as a consequence of particle-number and $\hat S_z$ conservation~\cite{Rissler2006,bogu15a,Stein2019}, so its four eigenvalues can be reconstructed without explicitly tracing out the many-body wavefunction. Denoting the spin-resolved 1-particle reduced density matrix (1-PDM) by $\gamma^{\sigma}_{pq} = \bra{\Psi}\hat{a}^{\dagger}_{p\sigma}\hat{a}_{q\sigma}\ket{\Psi}$ and the diagonal of the two-orbital particle-number correlator by $N^{\sigma\sigma'}_{pq} = \bra{\Psi}\hat{n}_{p\sigma}\hat{n}_{q\sigma'}\ket{\Psi}$, the on-site double occupation reads
\begin{equation}
    d_p \;=\; \langle \hat{n}_{p\uparrow}\hat{n}_{p\downarrow}\rangle
       \;=\; \tfrac{1}{2}\!\left(\langle \hat{n}_p\hat{n}_p\rangle
       - \langle \hat{n}_p\rangle\right),
\end{equation} where the second equality follows from $\hat{n}_{p\sigma}^2 = \hat{n}_{p\sigma}$ and $n_p = \gamma^{\uparrow}_{pp} + \gamma^{\downarrow}_{pp}$. The eigenvalues of $\rho^{p}$ then read~\cite{Rissler2006,Bogu2012,bogu15a}
\begin{equation}
    w_p^{(2)}          = d_p, \quad
    w_p^{(\uparrow)}   = \gamma^{\uparrow}_{pp} - d_p, \quad
    w_p^{(\downarrow)} = \gamma^{\downarrow}_{pp} - d_p, \quad
    w_p^{(0)}          = 1 - n_p + d_p,
\end{equation} and the single-orbital entropy follows directly as
\begin{equation}
    S_p \;=\; -\!\!\sum_{i\in\{0,\uparrow,\downarrow,2\}} w_p^{(i)} \ln w_p^{(i)}.
\end{equation}
This avoids constructing the full configurational projector in Equation~\ref{eqn:1o-rdm} and requires only quantities already produced by standard DMRG codes (e.g., \texttt{1pdm} and \texttt{1npc} in
\texttt{Block2}~\cite{zhai2021,zhai2023block2}). Although Eq.~\ref{eqn:1o-rdm} is most often encountered in the DMRG context, the one-particle density matrix $\gamma^{\sigma}_{pp}$ and the on-site double occupation $d_p = \langle \hat{n}_{p\uparrow}\hat{n}_{p\downarrow}\rangle$ are accessible from any correlated wavefunction for which the 1- and 2-RDMs (or, equivalently, the diagonal of the two-orbital number correlator) can be evaluated~\cite{bogu15a,Stein2019}. The formulation is also directly applicable to the unrestricted case: no assumption $\gamma^{\uparrow}_{pp} = \gamma^{\downarrow}_{pp}$ enters the derivation, so the two singly-occupied eigenvalues $w_p^{(\uparrow)} = \gamma^{\uparrow}_{pp} - d_p$ and $w_p^{(\downarrow)} = \gamma^{\downarrow}_{pp} - d_p$ remain distinct whenever the underlying state is spin-polarized.}

\section{Detailed AEGISS scaling}
Let $N_{AO}$ be the number of atomic orbitals (basis functions) in the whole system, $N_A$ the number of MOs kept for the active region $A$ after fragmenting/transforming, $N_E$ the number of entropy-preselected MOs (subset of $A$), $k$ the number of orbitals in the \textit{Large-CAS} treated by DMRG, $m$ the DMRG bond dimension, $s$ the number of DMRG sweeps, and $\sum_D N_D$ the total number of AOs included across all AO groups $D$. 

\noindent\textbf{Pre-processing (HF or KS-DFT)}:

\noindent For na\"ive HF or KS, the dominant cost per SCF iteration comes from the Fock build, scaling as $\mathcal{O}(N_{AO}^4)$, for the former, and $\mathcal{O}(N^3_{elec})$, for the latter. Any AO-to-MO transforms or orbital localization steps are typically $\mathcal{O}(N_{AO}^3)$.

\noindent\textbf{Single-orbital entropies via DMRG}:

\noindent Let $k$ denote the \textit{large-CAS} size in the approximate DMRG run. A two-site DMRG sweep has a cost of $\mathcal{O}(k m^3 + k^2 m^2)$, where $k\, m^3$ arises from local solutions and $k^2 m^2$ from Hamiltonian contractions. The total cost is $s \times \mathcal{O}(k m^3 + k^2 m^2)$. The entropy extraction itself (from 1o-RDMs) is $\mathcal{O}(k m^2)$-$\mathcal{O}(k m^3)$, negligible compared to sweep costs. This stage often dominates the total runtime when $k$ or $m$ are large.

\noindent\textbf{AO-projection}:

\noindent The AO-projection involves building the cross-integral overlap matrix $\mathbf{S}$ scales as $\mathcal{O}(N_{mAO}N_{AO})$.  For each AO group $D \in \{{\mathcal{D}}\}$, the projection $\mathbf{S}_D = \mathbf{P}_D \mathbf{S}$ is computed at $\mathcal{O}(N_D N_{AO})$ cost, followed by a matrix multiply $\mathbf{S}_D \mathbf{C}^A_E$ at $\mathcal{O}(N_D N_{AO} N_E)$. Weight computation and selection add $\mathcal{O}(N_D N_E)$ for summations. Summed over all groups $D$, the dominant projection cost is $\mathcal{O}((\sum_D N_D) N_{AO} N_E)$, much cheaper than DMRG.

\noindent\textbf{End-to-end scaling}:

\noindent With RI/DF, SCF pre-processing scales as $\mathcal{O}(N_{AO}^3)$. DMRG screening dominates for $k$ and high $m$, scaling as $s \times \mathcal{O}(k m^3 + k^2 m^2)$. The AO-projection is relatively inexpensive, scaling as $\mathcal{O}(N_{AO}^2)$ for overlap storage and $\mathcal{O}((\sum_D N_D) N_{AO} N_E)$ for projection and weighting. For large systems with small \textit{large-CAS}, SCF can dominate; for large \textit{large-CAS}, DMRG dominates.

\noindent\textbf{Practical optimization}:

\noindent To control scaling, it is advisable to use Resolution to Identity/Density-Fitting to keep SCF cubic, limit the Large-CAS to the size needed for reliable entropy estimates, and gradually increase $m$. Using symmetry-adapted or block-sparse DMRG can reduce effective prefactors.

\subsection{Post-selection screening}

\teal{Manual screening leverages chemical intuition and system-specific knowledge. Visual inspection remains a valuable tool to exclude Rydberg states or incorporate orbitals tailored to specific excitations of interest. When the molecular system exhibits point-group symmetry, irreducible representations provide an additional systematic criterion for discriminating which orbitals to include or exclude from the active space.}

\teal{Beyond manual screening, automated post-processing techniques can significantly enhance both the efficiency and interpretability of the active space.  Possible solution involves (Frozen) Natural Orbitals, (F)NOs,~\cite{SosaFNOs, taube2005frozen, taube2008, verma2021} which reduces the size of the correlated space by identifying and excluding virtual orbitals with negligible occupation—typically based on an MP2 or CCSD calculation—retaining accuracy while lowering computational cost. Similarly, localization of virtual orbitals improves chemical interpretability, reduces delocalization artifacts that complicate correlation treatment, and facilitates fragment-based embedding. Together, these refinements ensure that the active space remains both chemically meaningful and computationally tractable, particularly for large-scale systems as discussed in the following section.}

\section{Computational Details}
\label{sec:computational_details}

\teal{The atom-labeling pre-processing step is performed using the \texttt{MolTagger} application included in the AEGISS package, but in general, this procedure is software-agnostic and can equivalently be carried out with any molecular visualization tool (e.g., Avogadro~\cite{avogadro}, VESTA~\cite{vesta2008}) by appending the cluster identifier $\mathcal{C} \in \{\mathcal{R}\}$ to each atom.}

\teal{From the point of view of structure optimization instead, if required, we have used ORCA~\cite{ORCA} and Gaussian~\cite{g16}. All calculations following the pre-processing phase, i.e. building the initial guess MOs, have been performed using PySCF~\cite{pyscf, Sun2018, Sun2020}. As well as the calculations for validation of the selected active spaces are performed using the PySCF implementations of TD-DFT, CASCI, CASSCF, and NEVPT2.  When the identified active space exceeds the capacity of the PySCF FCI solver, DMRG-based solvers were used instead (DMRG-CI, DMRG-SCF, and where needed DMRG-NEVPT2), all available through the \texttt{dmrgscf} extension module, based on Block2~\cite{zhai2023block2} DMRG implementation.}

\teal{All NEVPT2 calculations were carried PySCF. After generating the orbital basis and the relative active-space, the specific subset of orbitals is extracted with \texttt{mcscf.CASSCF.sort\_mo}.  Mixed-spin state-averaged CASSCF was then performed using the \texttt{mcscf.state\_average\_mix\_} driver, which assigns separate FCI solvers to the singlet and triplet manifolds: both solvers are instances of \texttt{fci.direct\_spin1.FCI} wrapped by \texttt{fci.addons.fix\_spin} with a spin penalty of shift $0.2/0.5$, based on the system, and target $\langle S^2\rangle$ values of $0$ and $2$, respectively, ensuring spin-pure roots throughout the optimization. Uniform weights $w_i = 1/(n_s + n_t)$ were used across the $n_s$ singlets and $n_t$ triplets. 
After convergence of the SA-CASSCF orbitals (with \texttt{mc.canonicalization = True} and \texttt{mc.natorb = True}
producing canonical natural orbitals in the active space), separate \texttt{mcscf.CASCI} calculations were run for each multiplicity on top of the SA-CASSCF MOs.  The NEVPT2 correction was evaluated per root through \texttt{mrpt.NEVPT(mc, root=i).kernel()}, which implements the strongly contracted formulation of NEVPT2~\cite{angeli2001, angeli2001_2, angeli2002}.  Total energies were recovered as $E_i = E_i^{\text{CASCI}} + E_i^{\text{NEVPT2}}$, and states were sorted by CASCI energy to ensure a consistent reference across the full singlet/triplet manifold.  A per-root sanity check comparing the SA-CASSCF and CASCI energies confirmed that the diagonalization in each multiplicity block recovered the same roots present in the state-averaged calculation.} \teal{TD-DFT calculations for the Ru-based complexes were performed with ORCA~\cite{ORCA}.}

\teal{All DMRG calculations were performed with the Block2 package~\cite{zhai2023block2}. The single-orbital entropies can be extracted via \texttt{DMRGDriver.get\_orbital\_entropies()} if the python interface is used, or using Equation~\ref{eqn:1o-rdm} on the 1-RDM and correlation matrix, obtained from the from the resulting MPS.}

\teal{For the AO-projection step, all the relevant quantities and operators are computed using PySCF method.  AO labels are selected using \texttt{search\_ao\_labels($D$)}, and the non-orthogonal mixed-basis overlap matrix $[\mathbf{S}]_{\nu\mu}$ is computed via \texttt{Mole.intor('int1e\_cross')}.  A single AO label has the form ``$X_c\ \mathrm{AO}_s$'', where $X$ is the atom symbol, $c$ the cluster identifier, and $\mathrm{AO}_s$ the atomic orbital contribution, and labels can be specified individually or grouped.}

\teal{Once again, all steps of the active space selection methods, included the atom-tagging interactive application, as well as the visualization tools, and the full method are contained in AEGISS GitHub public repository.}

\subsection{\teal{Ferrocene}}
\label{apx:setup_ferrocene}
\teal{The ferrocene geometry was taken directly from the reference AVAS paper~\cite{Sayfutyarova_2017_AVAS}: two cyclopentadienyl (Cp) rings planar and parallel to the $z$-plane, optimized at the CCSD(T)/cc-pw-CVTZ level with $D_{5h}$ symmetry. Following the reference AVAS setup~\cite{Sayfutyarova_2017_AVAS}, the initial guess MOs were obtained from a canonical RHF calculation using the cc-pVDZ-DK basis with scalar relativistic correction X2C~\cite{x2c_1,x2c_2}. The resulting total number of MOs is $N = 542$.}

\teal{A reduced search space of $N_{\mathbf{A}} = 40$ MOs is defined. The approximate MPS is obtained from a DMRG calculation on CAS(40e,40o) using spin-restricted (SU2) symmetry, bond dimension up to 500 (starting from 100), and an FCIDUMP file generated by PySCF. Applying Eq.~\ref{eqn:s1} to the resulting MPS, the single-orbital entropies $S(1)_i$ are computed for each $\psi_i \in \{\psi_j\}_{j\in\mathbf{A}}$. Imposing the standard 10\% cutoff $\tau_E = 0.1$ on the relative entropies yields $N_E = 26$ entropy-screened MOs. For ferrocene, a single AO-label group $D = \text{``Fe 3d''}$ is used. The projection weight $w_p^D$ is computed for each entropy-screened orbital $\psi_p$, and all orbitals with $w_p^D > 0.5$ are selected for the final active space.}

\subsection{\teal{DOBNA}}
\label{apx:setup_dobna}
\teal{The $S_0$ geometry was taken directly from the supplementary information of Ref.~\cite{DOBNA_geom}, optimized at the B3LYP/6-31G(d) level with $C_2$ symmetry. Two sets of initial-guess MOs were generated.  The first set comes from a canonical RHF calculation with the 6-31G(d) basis.  Given the well-known delocalized character of canonical RHF orbitals, a second set was obtained from a correlated wavefunction: an MP2 calculation was run on top of the canonical RHF, the resulting 1-RDM was diagonalized, and eigenvectors reordered by occupation number to define a set of Natural Orbitals (NOs). The last setup, MP2-(F)NOs~\cite{MP2-NO}, the MP2 calculation is performed keeping frozen all the core orbitals that will be excluded from the active region when the entropy screening is performed. In other words, the difference between MP2-NOs and MP2-(F)NOs are the number of core orbitals used for the MP2 calculation. In all three setups, the total number of MOs is $N = 316$, with 70 occupied orbitals.}

\teal{In this case, we restricted in a subspace with size $N_{\mathbf{A}} = 40$. The approximate MPS, computed as for the ferrocene on a CAS(40e,40o),  is obtained using spin-restricted (SU2) symmetry, bond dimension up to 500 (starting from 100), and an FCIDUMP file generated by PySCF. Using Equation~\ref{eqn:s1}, we extracted all the SOE for each orbital in the active subspace, obtained maximum SOE, $S(1)_{\text{max}}$, 0.2540 and 0.2217, for RHF and MP2-NOs, respectively. For both the intial guesses, canonical RHF and MP2-NOs, we use two different thresholds, 10\% and 90\% of the maximum SOE in the subspace. With the looser threshold, we identified 18 active orbitals for RHF, while no orbital has been discarded for MP2-NOs, due to the higher correlation present in the active subspace. For the second threshold, the tighter one, in both cases, we identified 10 orbitals. After the entropy screening is completed, we then search for all the remaining MOs which AO-projection weight, $w_p^D$, on the composite label [''C/B 2pz"], is above 0.5. For the canonical RHF, for the loose threshold, all the 18 orbitals are above the AO-projection weight, as well as for the stricter one. For the MP2-NOs, the selected orbitals by the AO-projection step are 18, 9 occupied and 9 virtual, in the case of the looser threshold, while all the 10 pre-selected orbitals also pass the AO-selection, for the tighter threshold.}

\begin{figure}[!htb]
\centering

\begin{subfigure}[b]{0.48\linewidth}
    \centering
    \includegraphics[width=\linewidth,trim={0 4.5cm 0 2.7cm}, clip]{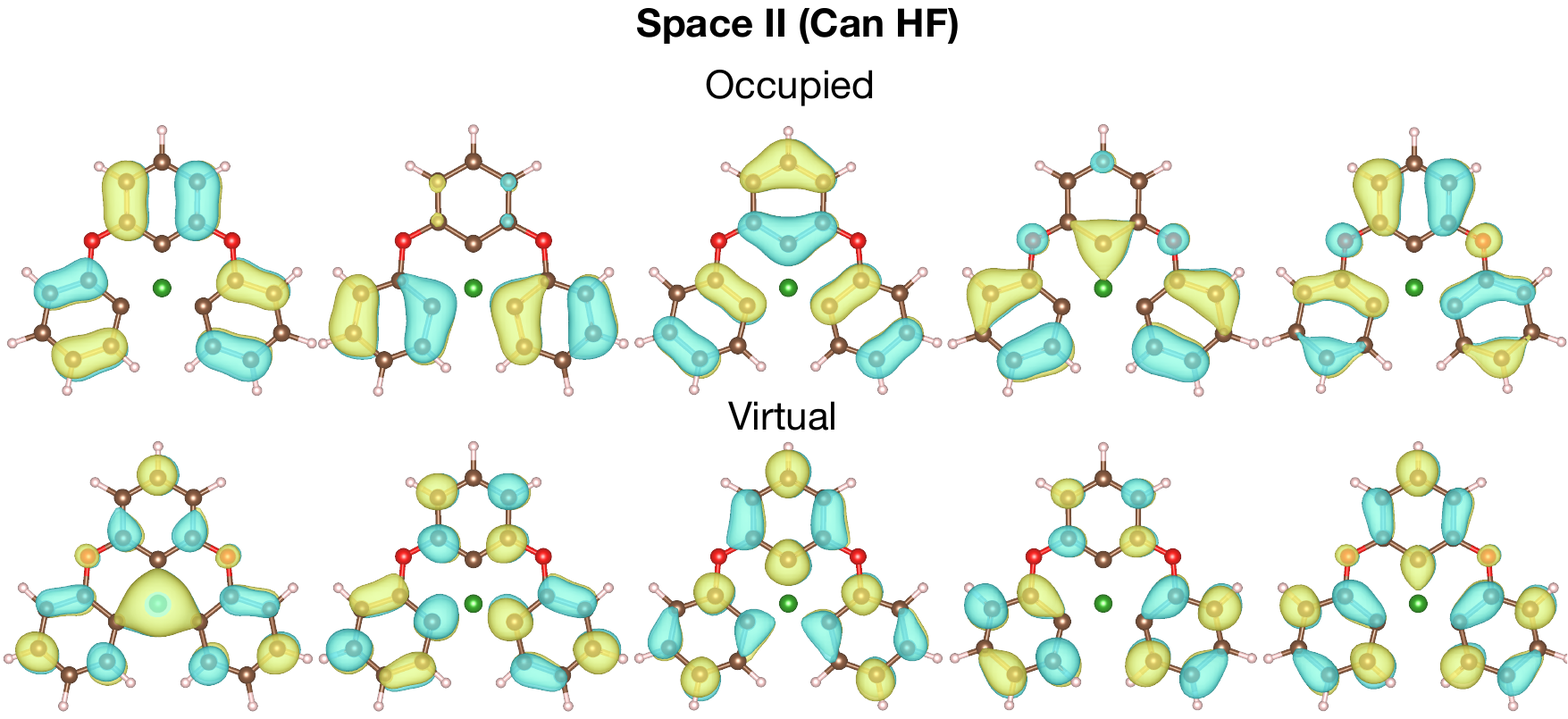}
    \caption{}
    \label{sfig:dobna_space_can}
\end{subfigure}
\hfill
\begin{subfigure}[b]{0.48\linewidth}
    \centering
    \includegraphics[width=\linewidth,trim={0 4.5cm 0 2.7cm}, clip]{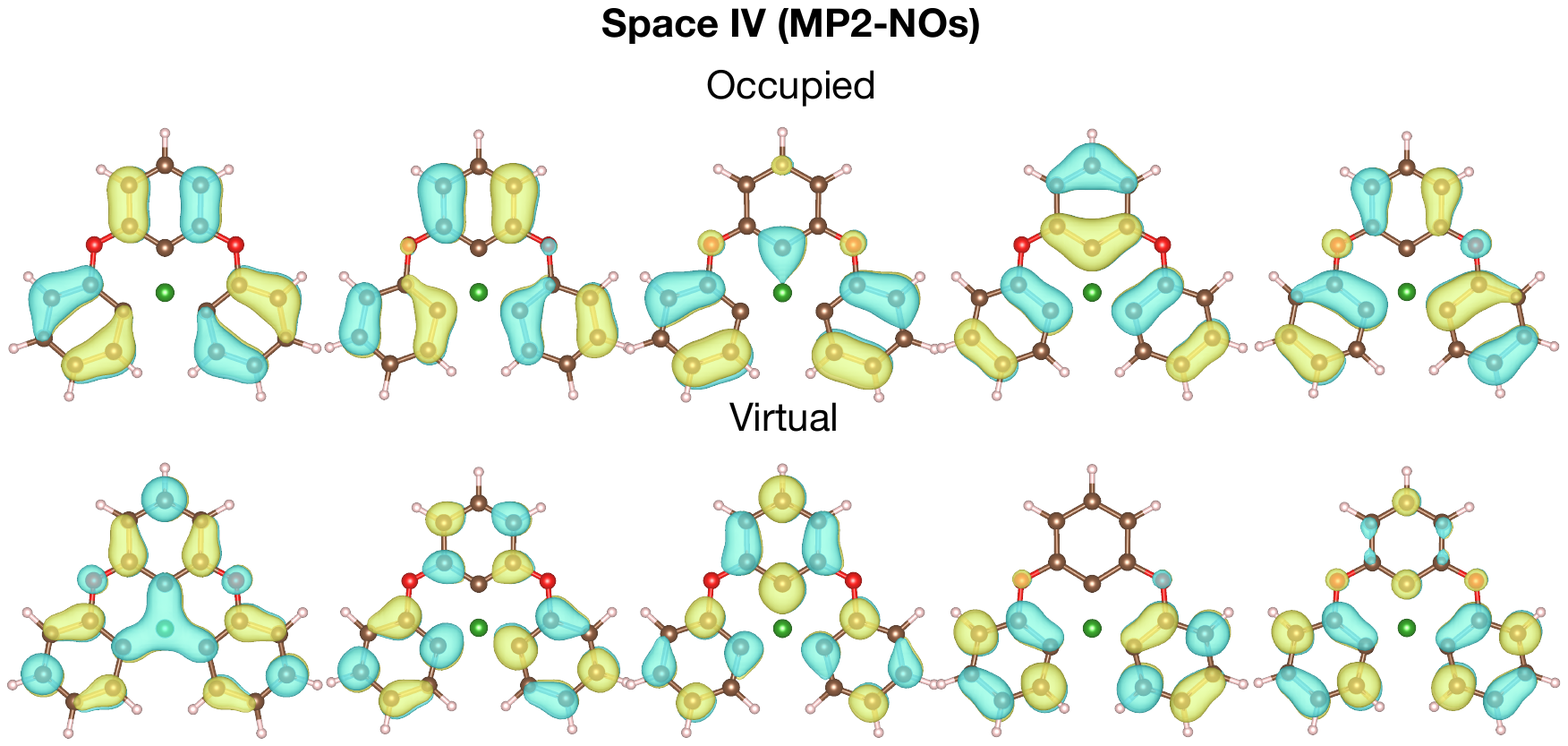}
    \caption{}
    \label{sfig:dobna_space_NOs}
\end{subfigure}
\caption{\teal{Final active spaces obtained by AEGISS on DOBNA using as starting guess Canonical Hartree-Fock orbitals\textbf{(a)} and MP2-natural orbitals~\cite{MP2-NO}\textbf{(b)}, with the 0.9 entropy screening threshold.}}
\label{fig:dobna_other_spaces}
\end{figure}

\subsection{\teal{Trans-Cl}}
\label{apx:setup_transcl}
The ground-state geometry was optimized in the gas phase with $C_{2v}$ symmetry using the B3LYP functional~\cite{B3LYP_1, B3LYP_2}, a 6-31G$^*$-type basis on all atoms except Ru, and a def2-TZVP basis~\cite{def2tzvp} with the quasi-relativistic Stuttgart–Dresden pseudo-potential~\cite{pp_transcl} for Ru.  The optimization was performed with ORCA~\cite{ORCA}, with settings chosen to match the reference calculations of Ref.~\cite{TransCl} as closely as possible. The correct $C_{2v}$ symmetry is enforced in both ORCA and PySCF by raising the symmetry threshold during the mean-field step.  For the RHF, CASSCF, and DMRG calculations, def2-TZVP was used on all atoms, while for Ru(II), def2-TZVP with the quasi-relativistic Stuttgart–Dresden pseudo-potential was employed, which corresponds to cc-pVDZ-pp in PySCF. This basis set and effective core potential is consistent with the geometry optimization setup.  (The basis used in the reference work is ANO-RCC-VTZP~\cite{ano_rcc_tzvp_1, ano_rcc_tvzp_2}.)  \teal{The full molecular orbital space comprises $N = 658$ MOs, with the HOMO sitting at MO index 80, counting for 160 electrons in the system.} 
\teal{A reduced search space is defined by removing core orbitals and high-energy virtual MOs, yielding $N_{\mathrm{LCAS}} = 90$ MOs with 32 orbitals frozen as inactive core.  The DMRG calculation is thus set up as CAS(92,90).  The approximate MPS is obtained with fixed bond dimension 500 in a non-spin-adapted (SZ) framework.} Once the approximate MPS is available, $S(1)_i$ is computed for each MO via Eq.~\ref{eqn:s1} and the entropy pre-selection is applied.  The maximum single orbital entropy is $S(1)_{\max} = 0.226$. The standard 10\% cutoff reduces the pool of 90 orbitals to $N_E = 42$ pre-selected MOs.

The AO labels of interest for each cluster $\mathcal{C}_i$ are the five individual Ru 4d contributions (``4d$_{xz}$'', ``4d$_{yz}$'', ``4d$_{xy}$'', ``4d$_{x^2-y^2}$'', ``4d$_{z^2}$'') and, for recovering $\pi/\pi^*$ orbitals, the 2p$_z$ AOs of the labelled carbons and nitrogens on the ligands, excluding contributions from the CO moieties.  

Reference calculations were obtained from PySCF state-averaged CASSCF. 

\subsection{\teal{TLD-1433}}
\label{apx:tld1433_setup}
For the TLD-1433 complex, the reference paper~\cite{Alberto2016} considers geometries differing in the orientation of the thiophene rings: either with all S atoms pointing in the same direction, or alternating.  We found the alternating configuration to be more stable.  Additionally, the orientation of the H atom on the imidazole ring relative to the first thiophene S atom gives rise to two further variants.  Only the same-direction H–S arrangement was studied in the reference work~\cite{Alberto2016, McFarland2019}.  Here, a new local minimum was identified for the alternated-ring / opposite H–S configuration.  Ground-state geometries were optimised with Gaussian~\cite{g16} using PBE0~\cite{PBE0_1, PBE0_2}, def2-TZVP, and a Polarizable Continuum Model (PCM)~\cite{TRUONG1995253, Barone1998} for water as solvent.  Geometry energies and full analysis are reported in Section~E. 
RHF calculations were performed with PySCF using the def2-SVDP basis~\cite{def2svpd} with the cc-pVDZ-pp pseudo-potential for Ru(II). Solvent effects in water were included via the PySCF implementation of the Conductor-PCM (C-PCM)~\cite{TRUONG1995253, Barone1998}.  The full molecular orbital space comprises $N_{\mathbf{A}} = 1559$ MOs; the HOMO is MO number 225 with a total of 500 electrons.

Working with the full orbital space is intractable, so a reduced search space is defined by removing core orbitals and high-energy virtual MOs, yielding $N_{\mathrm{LCAS}} = 100$ MOs with 185 orbitals frozen as inactive core.  The DMRG calculation is set up as CAS(80,100).  Once the approximate DMRG wavefunction is obtained, $S(1)_p$ is computed for each MO in $\{\psi_p^{\mathrm{LCAS}}\}$ via Eq.~\ref{eqn:s1}.  The maximum single-orbital entropy is $S(1)_{\max} = 0.201$, giving $\tau_E = 0.0201$.  The entropy-based selection reduces the pool of 100 MOs to $N_E = 51$ pre-selected MOs.

The reference calculation for TLD-1433 is obtained from DMRG-CI using Block2. The system exhibits a closed-shell singlet ground state, so the highest available symmetry (\texttt{SymmetryTypes.SU2}) is used, with bond dimension \texttt{m = 1000}, \texttt{n\_sweeps = 100}, and noise schedule \texttt{noises = [1e-4] * n\_sweeps//2 + [1e-5] * n\_sweeps//2 + [0]}. \teal{The singlet states were obtained through a state-averaged DMRG-CI calculation targeting two roots, followed by two independent state-specific DMRG-CI refinements. The triplet state was instead computed through a single DMRG-CI calculation with target \texttt{spin=2}. Once the matrix product state (MPS) had converged, the solvent correction was evaluated by contracting the one-particle reduced density matrix (1-RDM) with the solvent potential.}

\newpage
\section{DOBNA Benchmarks}
\label{apx:dobna_benchmarks}

\begin{table}[!htb]
\centering
\small
\begin{tabular}{c|ccc}
\toprule
& \multicolumn{3}{c}{CASSCF} \\
Space & $S_1$ & $T_1$ & $\Delta E_{\text{S-T}}$ \\\hline\hline
I & $\sim$ & $\sim$ & $\sim$ \\
II &  4.769 & 3.712 & 1.057 \\
III & 4.621	& 3.884& 0.737 \\
IV &  4.768 & 3.712 & 1.056\\
V & 4.415 & 3.832 & 0.583 \\
VI & 5.395 & 4.198 & 1.198  \\
\bottomrule
\end{tabular}
\caption{\teal{DOBNA vertical excitation energies and singlet-triplet gap for the six SA-CASSCF. State-averaged on 2 singlets and 2 given the computational cost of the large spaces. All energies in eV. Active space I did not converge within the max number of SCF cycles imposed, i.e. 50.}}
\label{tab:dobna_6_spaces}
\end{table}

\begin{table}[!htb]
    \centering
    \begin{tabular}{c|cc|cc|cc}
    \toprule
         ~ &  \multicolumn{2}{c|}{PBE}& \multicolumn{2}{c|}{B3LYP} & \multicolumn{2}{c}{Cam-B3LYP} \\
    Root & $\Delta E$ & $f$& $\Delta E$ & $f$& $\Delta E$ & $f$ \\ \hline\hline
    S$_1$ & 2.943 & 0.078 & 3.427 & 0.121 & 3.949 &  0.197\\
    S$_2$ & 3.639 & 0.010 & 4.186 & 0.033 & 4.728 &  0.066\\
    S$_3$ & 3.987 & 0.000 & 4.527 & 0.019 & 4.938 &  0.027\\
    S$_4$ & 4.090 & 0.013 & 4.534 & 0.017 &4.947 &  0.009\\
    S$_5$ & 4.133 & 0.036 & 4.727 & 0.120 &5.210 &  0.103\\
    \bottomrule
    \end{tabular}
    \caption{\teal{Electronic transition energies (in eV) of the first 5 singlet states from TD-DFT calculation performed on DOBNA in GAS PHASE. Functionals: PBE, B3LYP and Cam-B3LYP. Basis-set used 6-31G(d) for all the atoms. Additional data can be found in Dos Santos et al.~\cite{review_TADF}}}
    \label{tab:dobna_singlets_tddft}
\end{table}
\begin{table}[!htb]
    \centering
    \begin{tabular}{c|c|c|c}
    \toprule
    Root & PBE & B3LYP &Cam-B3LYP  \\ \hline\hline
    T$_1$ & 2.550 & 2.903 &  3.206 \\
    T$_2$ & 3.162 & 3.352 &  3.274 \\
    T$_3$ & 3.447 & 3.456 &  3.356 \\
    T$_4$ & 3.475 & 3.611 & 3.644 \\
    T$_5$ & 3.707 & 3.901 & 3.889 \\
    \bottomrule
    \end{tabular}
    \caption{\teal{Electronic transition energies (in eV) of the first 5 triplet states from TD-DFT calculation performed on DOBNA in GAS PHASE. Functionals: PBE, B3LYP and Cam-B3LYP. Basis-set used 6-31G(d) for all the atoms. Additional data can be found in Dos Santos et al.~\cite{review_TADF}}}
    \label{tab:dobna_triplets_tddft}
\end{table}
\newpage

\section{Trans-Cl Benchmarks}
\label{apx:transcl_benchmarks}
\subsection{State-Average CASSCF}
\begin{table}[!htb]
\centering
\setlength\tabcolsep{4pt}
\begin{tabular}{c|cc|cc}
\toprule
 & \multicolumn{2}{c|}{I} & \multicolumn{2}{c}{II} \\
Root & S$_n$ & T$_n$ & S$_n$ & T$_n$ \\\hline
1  & 5.378 & 3.819 & 3.463 & 2.702 \\
2  & 5.828 & 4.848 & 3.509 & 2.802 \\
3  & 5.853 & 4.933 & 5.060 & 3.899 \\
4  & 6.614 & 5.260 & 5.264 & 3.929 \\
5  & 7.028 & 5.318 & 5.374 & 4.180 \\
6  & 7.862 & 5.364 & 5.449 & 4.457 \\
7  & 7.887 & 5.533 & 5.928 & 4.653 \\
8  & 7.947 & 6.627 & 6.199 & 4.942 \\
9  & 8.304 & 6.756 & 6.609 & 5.631 \\
10 & 8.538 & 7.879 & 6.678 & 5.962 \\
11 & 8.690 & 8.258 & 7.118 & 5.970 \\\bottomrule
\end{tabular}
\caption{\teal{Singlet ($\text{S}_n$) and triplet ($\text{T}_n$) excited state transition energies (in eV) relative to $\text{S}_0$, obtained from SA-CASSCF calculations on the Trans-Cl complex with active spaces I and II. Computational details reported in Section~\ref{apx:setup_transcl}.}}
\label{tab:transcl_casscf_full}
\end{table}

\subsection{TDDFT}
The performance of several TD-DFT flavors has been assessed as we need a general method which allows describing the different transitions contributing to the UV-vis spectrum. While it is known that solvent effects are found to be mandatory to obtain spectroscopic accuracy, especially in the case of MLCT states, for the investigation of this prototype complex we only performed gas phase calculations to the purpose of benchmarking TDDFT calculations against CASSCF calculations based on the active spaces we found. The results of the TD-DFT calculations are collected in Table \ref{tab:transcl_tddft_functionals_s} for the singlet states, with also the oscillator strengths, and in Table \ref{tab:transcl_tddft_functionals_t} for the triplet states.

\teal{The TD-DFT excitation energies obtained here reproduce the main features of the reference \cite{TransCl} spectrum. The MLCT manifold is captured by long-range-corrected and high-exact-exchange functionals (CAM-B3LYP, M06-2X) but is systematically red-shifted by GGA and mild-hybrid functionals (PBE0, B3LYP), whereas the MC band is robust across functionals. This is consistent with the established behaviour of TD-DFT for Ru(II) polypyridyl excitations.} 
As a result, we expect our findings to be aligned with the description that MC transitions are rather robust to any of the functionals tested while MLCT states are only well described with functionals bearing intermediate amounts of exact exchange, i.e. PBE0 and B3LYP, possibly in combination with solvent effects. IL states are also best described with these functionals. 

\begin{table}[!htb]
\fontsize{8pt}{10pt}\selectfont
\centering
    \setlength\tabcolsep{1pt}
\begin{tabular}{>{\centering\hspace{0pt}}m{0.1\linewidth}||>{\centering\hspace{0pt}}m{0.125\linewidth}>{\centering\hspace{0pt}}m{0.083\linewidth}|>{\centering\hspace{0pt}}m{0.125\linewidth}>{\centering\hspace{0pt}}m{0.083\linewidth}|>{\centering\hspace{0pt}}m{0.125\linewidth}>{\centering\hspace{0pt}}m{0.083\linewidth}|>{\centering\hspace{0pt}}m{0.125\linewidth}>{\centering\arraybackslash\hspace{0pt}}m{0.083\linewidth}}
 & \multicolumn{2}{>{\centering\hspace{0pt}}m{0.208\linewidth}|}{PBE} & \multicolumn{2}{>{\centering\hspace{0pt}}m{0.208\linewidth}|}{PBE0} & \multicolumn{2}{>{\centering\hspace{0pt}}m{0.208\linewidth}|}{B3LYP} & \multicolumn{2}{>{\centering\arraybackslash\hspace{0pt}}m{0.208\linewidth}}{CAM-B3LYP} \\
State & $\Delta E$ & $f$ & $\Delta E$ & $f$ & $\Delta E$ & $f$ & $\Delta E$ & $f$ \\ 
\hline\hline
$S_1$ & 1.395 & \textit{0.0003} & 2.471 & \textit{0.0003} & 2.282 & \textit{0.0003} & 3.303 & \textit{0.0002} \\
$S_2$ & 1.56 & \textit{\textbf{0.0122}} & 2.58 & \textit{\textbf{0.0141}} & 2.40 & \textit{\textbf{0.0132}} & 3.38 & \textit{0.0000} \\
$S_3$ & 2.164 & \textit{0.0028} & 3.301 & \textit{0.0000} & 3.225 & \textit{0.0025} & 3.41 & \textit{\textbf{0.0191}} \\
$S_4$ & 2.265 & \textit{0.0008} & 3.377 & \textit{0.0031} & 3.252 & \textit{0.0000} & 3.472 & \textit{0.0038} \\
$S_5$ & 2.375 & \textit{0.0000} & 3.466 & \textit{0.0024} & 3.284 & \textit{0.0002} & 4.131 & \textit{0.0041} \\
$S_6$ & 2.386 & \textit{0.0000} & 3.529 & \textit{0.0007} & 3.306 & \textit{0.0012} & 4.292 & \textit{0.0000} \\
$S_7$ & 2.394 & \textit{0.0105} & 3.538 & \textit{0.0000} & 3.139 & \textit{0.0000} & 4.41 & \textit{0.0007} \\
$S_8$ & 2.469 & \textit{0.0044} & 3.547 & \textit{0.0014} & 3.347 & \textit{0.0028} & 4.461 & \textit{0.0000} \\
$S_9$ & 3.078 & \textit{0.0000} & 3.683 & \textit{0.0084} & 3.439 & \textit{0.0083} & 4.49 & \textit{0.0008} \\
$S_{10}$ & 3.137 & \textit{0.0016} & 3.750 & \textit{0.0025} & 3.506 & \textit{0.0027} & 4.582 & \textit{0.0021} \\
$S_{11}$ & 3.149 & \textit{0.0005} & 3.887 & \textit{0.0048} & 3.791 & \textit{0.0045} & 4.673 & \textit{0.0079}
\end{tabular}
\captionof{table}{Electronic transition energies (in eV) and oscillator strength of the first 11 singlets states from TD-DFT calculation performed on Trans-Cl complex in GAS PHASE. Functionals: PBE, PBE0, B3LYP, and CAM-B3LYP. Basis-set used 6-311G* for all the atoms except for Ruthenium. On Ruthenium def2-TZVP basis has been used with quasi-relativistic Stuttgart-Dresden pseudopotential.}
\label{tab:transcl_tddft_functionals_s}
    \end{table}
\hfill
\begin{table}[!htb]
\fontsize{8pt}{10pt}\selectfont
\centering
\begin{tabular}{c||c|c|c|c}

State & PBE & PBE0 & B3LYP & CAM-B3LYP \\ 
\hline\hline
$T_1$ & 1.368 & 2.439 & 2.254 & 2.847 \\
$T_2$ & 1.453 & 2.509 & 2.324 & 2.981 \\
$T_3$ & 2.144 & 2.753 & 2.774 & 3.212 \\
$T_4$ & 2.233 & 2.875 & 2.888 & 3.327 \\
$T_5$ & 2.347 & 3.271 & 3.207 & 3.429 \\
$T_6$ & 2.353 & 3.447 & 3.271 & 3.633 \\
$T_7$ & 2.364 & 3.458 & 3.277 & 3.832 \\
$T_8$ & 2.439 & 3.495 & 3.278 & 4.207 \\
$T_9$ & 2.636 & 3.503 & 3.285 & 4.231 \\
$T_{10}$ & 2.741 & 3.512 & 3.409 & 4.380\\
$T_{11}$ & 3.034 & 3.651 & 3.440 & 4.394
\end{tabular}

    \captionof{table}{Electronic transition energies (in eV) of the first 11 triplet states from TD-DFT calculation performed on Trans-Cl complex in GAS PHASE. Functionals: PBE, PBE0, B3LYP and CAM-B3LYP. Basis-set used 6-311G* for all the atoms except for Ruthenium. On Ruthenium def2-TZVP basis has been used with quasi-relativistic Stuttgart-Dresden pseudopotential.}
    \label{tab:transcl_tddft_functionals_t}
\end{table}

\begin{figure}[!htb]
    \centering
    \includegraphics[width=0.7\linewidth]{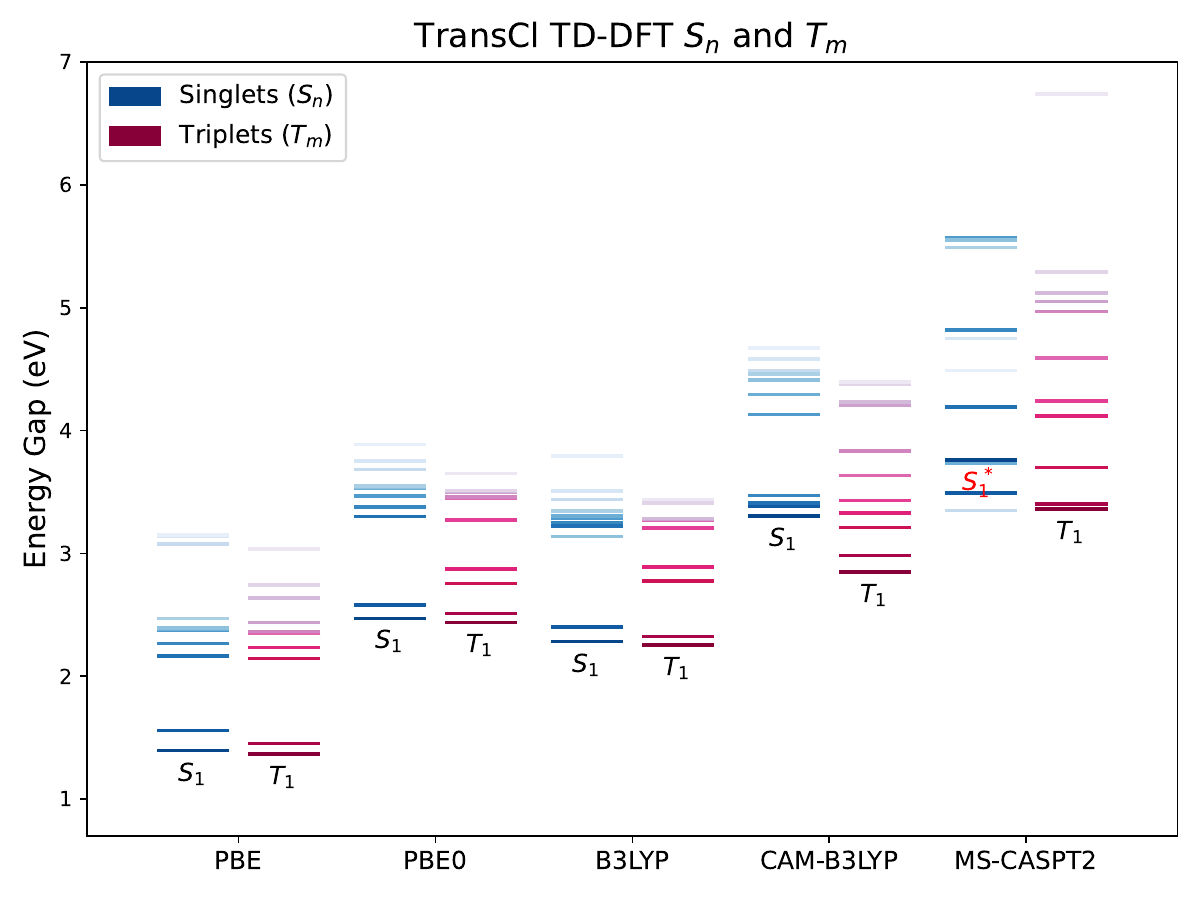}
    \caption{Diagram of $S_n$ and $T_m$ vertical excitation states for Trans-Cl complex obtained from TD-DFT calculation with different functionals (PBE, PBE0, B3LYP, and CAM-B3LYP) and reference benchmark MS-CASPT2. Singlet states computed by MS-CASPT2 are subject to a state reordering.  }
    \label{fig:transcl_tddft_ve}
\end{figure} 

Finally, as highlighted in Figure \ref{fig:transcl_tddft_ve}, it is clear that TDDFT alone provides a strongly functional-dependent description of the excitation spectra and great caution should be used when employing this tool for predicting UV-vis spectra.

\newpage
\section{TLD1433 Benchmarks}
\label{si_sec:TLD1433_benchmark}
In our study, we optimized the ground-state geometry for the conformation with alternated S-Ring and opposite H-S-Position, namely TLD1433-3(G), obtaining a new local minima of the geometry in both gas and solvent.
Our structure, TLD1433-3(G), is lower in energy with respect to the optimal structure of the reference paper, TLD1433-4(G), of 0.042 mHa in gas phase and 1.84 mHa in $H_2O$.

\subsection{TDDFT}
In reference \cite{Alberto2016}, a series of preliminary calculations using various exchange-correlation density functionals on smaller Ru compounds in methanol, with available experimental data, showed that PBE0 accurately reproduces the geometrical parameters. Therefore, we selected it as the most suitable XC functional for ground-state molecular optimizations. However, the computed absorption spectra indicate that M06 significantly outperforms the other XC functionals, particularly at longer wavelengths. The excellent performance of M06 has also been noted in previous studies on transition-metal complexes. To evaluate the reliability of our data, we focused on comparing our results with those of Alberto et al\cite{Alberto2016}. In their study, they presented M06 DFT results for TDL1433, and we find that our results are in excellent agreement with theirs. However, in this work, in particular to gain a deeper understanding of the excitations of this molecular system as well as to be able to accurately select the active space for our quantum circuit simulations, we decided to go beyond existing literature calculations. Specifically, we explore the effect of different exchange-correlation functionals on the excitations, including M062x (27$\%$ exact HF exchange), which has twice the exact HF exchange of M06 (15$\%$ exact HF exchange).

In previous systems, we have consistently tested and reported results both with and without the TD approximation. For this particular case, we have also run calculations with both options. However, to simplify the discussion and given the limited experimental data available to definitively confirm the accuracy of one approach over the other, we find that using the TD approximation is not only acceptable but also offers the advantage of faster computation at this stage. This choice is based on the practical consideration that the TD approximation simplifies the calculations without introducing significant discrepancies, allowing us to make reasonable progress while acknowledging the limitations posed by the lack of sufficient experimental validation. We also provide a table to compare the results for both approaches. Still, from now on, we will primarily present results based on the TDA-based approximation. Furthermore, the main focus and results of current benchmark investigation is on calculations with implicit solvent in water.

\begin{table}[!htb]
\fontsize{7pt}{10pt}\selectfont
\centering
    \setlength\tabcolsep{1pt}
\begin{tabular}{>{\centering\hspace{0pt}}m{0.1\linewidth}||>{\centering\hspace{0pt}}m{0.123\linewidth}>{\centering\hspace{0pt}}m{0.083\linewidth}|>{\centering\hspace{0pt}}m{0.121\linewidth}>{\centering\hspace{0pt}}m{0.081\linewidth}|>{\centering\hspace{0pt}}m{0.121\linewidth}>{\centering\hspace{0pt}}m{0.081\linewidth}|>{\centering\hspace{0pt}}m{0.121\linewidth}>{\centering\arraybackslash\hspace{0pt}}m{0.081\linewidth}}
 & \multicolumn{2}{>{\centering\hspace{0pt}}m{0.206\linewidth}|}{PBE0} & \multicolumn{2}{>{\centering\hspace{0pt}}m{0.202\linewidth}|}{B3LYP} & \multicolumn{2}{>{\centering\hspace{0pt}}m{0.202\linewidth}|}{CAM-B3LYP} & \multicolumn{2}{>{\centering\arraybackslash\hspace{0pt}}m{0.202\linewidth}}{M06} \\
State & $\Delta E$ & $f$ & $\Delta E$ & $f$ & $\Delta E$ & $f$ & $\Delta E$ & $f$  \\
\hline\hline
$S_1$ & 2.625  & \textit{0.0008} & 2.473& \textit{0.0011} & 3.133 & \textit{\textbf{1.9516}} & 2.387 & \textit{0.0008}\\
$S_2$ & 2.709  & \textit{0.0017} & 2.542& \textit{0.0013} & 3.134 & \textit{0.3315} & 2.462 & \textit{0.0016} \\
$S_3$ & 2.718  & \textit{0.0334} & 2.563& \textit{0.0597} & 3.215 & \textit{0.0141} & 2.467 & \textit{0.0006} \\
$S_4$ & 2.733  & \textit{\textbf{0.9679}} & 2.569& \textit{\textbf{0.4472}} & 3.249 & \textit{0.0014} & 2.569 & \textit{0.0042}\\
$S_5$ & 2.798  & \textit{0.0080} & 2.658& \textit{0.0102} & 3.300 & \textit{0.0011} & 2.620 & \textit{\textbf{0.4991}}\\
$S_6$ & 2.873  & \textit{0.0093} & 2.717& \textit{0.0073} & 3.352 & \textit{0.0299} & 2.639 & \textit{0.0191}\\
$S_7$ & 2.886  & \textit{0.1651} & 2.720& \textit{0.0122} & 3.388 & \textit{0.0121} & 2.705 & \textit{0.4908}\\
$S_8$ & 2.921  & \textit{1.2190} & 2.764& \textit{1.2701} & 3.417 & \textit{0.0007} & 2.780 & \textit{0.1848}\\
$S_9$ & 2.995  & \textit{0.1885} & 2.816& \textit{0.4539} & 3.429 & \textit{0.2406} & 2.868 & \textit{0.8037} \\
$S_{10}$ & 2.998  & \textit{0.0298} & 2.831 & \textit{0.1042} & 3.674 & \textit{0.0461} & 2.895 & \textit{0.3348}\\
$S_{11}$ & 3.066  & \textit{ 0.0027} & 2.874 & \textit{0.0500} & 3.805 & \textit{0.0032} & 2.941 & \textit{0.1182}\\
$S_{12}$ & 3.138  & \textit{ 0.0175} & 2.907 & \textit{0.0995} & 3.956 & \textit{0.0568} & 3.076 & \textit{0.0023}\\
$S_{13}$ & 3.164  & \textit{ 0.0059} & 2.972 & \textit{0.0062} & 4.009 & \textit{0.0030} & 3.112 & \textit{0.0031} \\
$S_{14}$ & 3.288  & \textit{ 0.0532} & 3.081& \textit{0.0932} & 4.071 & \textit{0.0777} & 3.153 & \textit{0.0142}\\
$S_{15}$ & 3.327  & \textit{ 0.0137} & 3.134 & \textit{0.0151} & 4.101 & \textit{0.0003} & 3.161 & \textit{0.0086} 
\end{tabular}
    \captionof{table}{Electronic transition energies (in eV) and oscillator strength of the first 15 singlet states from TD-DFT calculation performed on TLD1433 complex in implicit solvent WATER with TDA approximation. Functionals: PBE0, B3LYP, CAM-B3LYP, and M06. Basis-set used def2-TZVP for all the atom and quasi-relativistic Stuttgart-Dresden pseudopotential on Ruthenium.}
    \label{tab:TLD1433_tddft_functionals_tdaW_s}
\end{table}

\begin{table}[!htb]
\fontsize{8pt}{10pt}\selectfont
\centering
\begin{tabular}{c||c|c|c|c}
State & M062x W & $f$ & M062x G & $f$ \\
\hline\hline
$S_1$ & 3.182 &  \textit{2.3199} & 2.278 &  \textit{0.0442} \\
$S_2$ & 3.381 & \textit{0.0011} & 2.455 & \textit{0.0005}\\
$S_3$ & 3.442 & \textit{0.0021} & 2.517 & \textit{0.0125}\\
$S_4$ & 3.513 & \textit{0.0058} & 2.663 & \textit{0.3428}\\
$S_5$ & 3.521 & \textit{0.0105}& 3.048 & \textit{0.6417}\\
$S_6$ & 3.570 & \textit{0.0132} & 3.313 & \textit{ 0.1432}\\
$S_7$ & 3.613 & \textit{0.0077}& 3.372 & \textit{0.0043}\\
$S_8$ & 3.621 & \textit{0.0023} & 3.428 & \textit{0.0013}\\
$S_9$ & 3.657 & \textit{0.0124} & 3.445 & \textit{0.0006}\\
$S_{10}$ & 3.686 & \textit{0.0152} & 3.544 & \textit{0.0019}\\
$S_{11}$ & 3.706 & \textit{0.1835} & 3.588 & \textit{0.0026}\\
$S_{12}$ & 3.712 & \textit{0.0137} & 3.600 & \textit{0.0002}\\
$S_{13}$ & 3.795 & \textit{0.0916} & 3.610 & \textit{0.0017} \\
$S_{14}$ & 3.798 & \textit{0.0052} & 3.652 & \textit{0.0009}\\
$S_{15}$ & 4.006 & \textit{0.0426} & 3.661 & \textit{0.0017}
\end{tabular}
    \captionof{table}{Electronic transition energies (in eV) of the first 15 singlet states from TD-DFT calculation performed on TLD1433 complex in implicit solvent WATER (W) and GAS (G) phase with TDA approximation. Functionals M062x. Basis-set used def2-TZVP for all the atom and quasi-relativistic Stuttgart-Dresden pseudopotential on Ruthenium.}
    \label{tab:TLD1433_tddft_functionals_tdaW_s2x}
\end{table}

\begin{table}[!htb]
\fontsize{8pt}{10pt}\selectfont
\centering
\begin{tabular}{c||c|c|c|c|c}
State & PBE0 & B3LYP & CAM-B3LYP & M06 & M062x \\
\hline\hline
$T_1$ & 2.007 & 1.974 & 2.141 & 2.059 & 2.349\\
$T_2$ & 2.464 & 2.344 & 2.820 & 2.274 & 3.037\\
$T_3$ & 2.465 & 2.361 & 2.850 & 2.339 & 3.092\\
$T_4$ & 2.549 & 2.419 & 2.911 & 2.345 & 3.114\\
$T_5$ & 2.580 & 2.435 & 2.965 & 2.369 & 3.151\\
$T_6$ & 2.617 & 2.492 & 3.011 & 2.439 & 3.243 \\
$T_7$ & 2.651 & 2.533 & 3.033 & 2.463 & 3.307\\
$T_8$ & 2.694 & 2.569 & 3.076 & 2.509 & 3.318\\
$T_9$ & 2.722 & 2.583 & 3.192 & 2.577 & 3.344\\
$T_{10}$ & 2.749 & 2.645 & 3.269 & 2.588 & 3.355\\
$T_{11}$ & 2.802 & 2.653 & 3.280 & 2.690 & 3.375\\
$T_{12}$ & 2.818 & 2.676 & 3.327 & 2.720 &3.431\\
$T_{13}$ & 2.914 & 2.796 & 3.356 & 2.884 & 3.465\\
$T_{14}$ & 3.047 & 2.832 & 3.539 & 2.896 & 3.498\\
$T_{15}$ & 3.110 & 2.868 & 3.546 & 3.071 & 3.544
\end{tabular}
    \captionof{table}{Electronic transition energies (in eV) of the first 15 triplet states from TD-DFT calculation performed on TLD1433 complex in implicit solvent WATER with TDA approximation. Functionals: PBE0, B3LYP, CAM-B3LYP, and M062x. Basis-set used def2-TZVP for all the atom and quasi-relativistic Stuttgart-Dresden pseudopotential on Ruthenium.}
    \label{tab:TLD1433_tddft_functionals_tdaW_t}
\end{table}

 We report electronic transition energies and oscillator strength for singlet and triplet states for calculations in aqueous solution are reported in Tables \ref{tab:TLD1433_tddft_functionals_tdaW_s}, \ref{tab:TLD1433_tddft_functionals_tdaW_s2x} and \ref{tab:TLD1433_tddft_functionals_tdaW_t}, respectively. 

\begin{figure}[tbhp]
\centering
\includegraphics[scale=0.5]{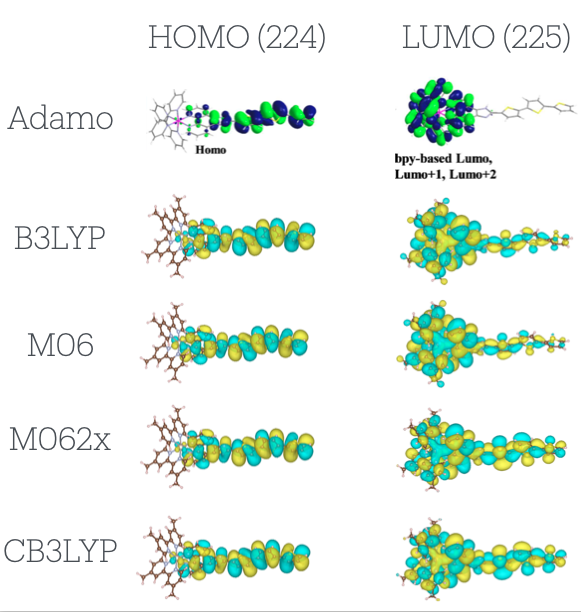}
\caption{HOMO and LUMO orbitals as computed by DFT and compared to Ref.~\cite{Alberto2016}.} 
\label{fig:hl_orbs}
\end{figure}

\begin{figure}[!tbhp]
\centering
\includegraphics[scale=0.5]{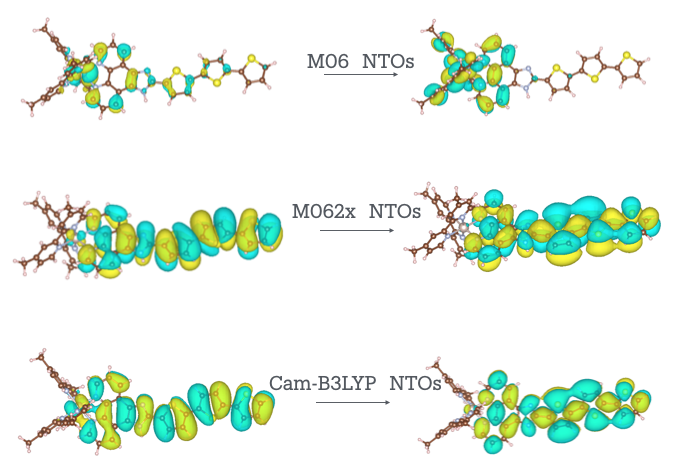}
\caption{Natural Transition Orbitals for S$_1$ computed with PBE0, M06, CAM-B3LYP and M062x.} 
\label{fig:ntos}
\end{figure}

Let us start from inspecting the character of the first bright singlet states (higher oscillator strengths) for all the functionals. We find that not all functionals predict the first singlet state to be the first bright state but all functionals, except M06, show that the first bright singlet has a stronger HOMO-LUMO character (cf Figure \ref{fig:ntos}). As shown in Figure \ref{fig:hl_orbs}, DFT consistently predicts HOMO to be solely localised on the tail while LUMO has both contributions from the ligands around the metal and smaller contribution on the tail, suggesting this excitation can be classified as ligand to ligand. To gain a deeper understanding of the character of the excitation we look into the natural transition orbitals (NTOs) for the first bright state. In Figure \ref{fig:ntos} we report the NTOs for the bright transition singlet and most relevant contributions (for simplicity we are only showing M06, CAM-B3LYP and M062x, but B3LYP shows same behaviour as CAM-B3LYP and M062x). We observe that, focusing only on M06  (see also Ref.~\cite{Alberto2016}), gives a limited picture of this excitation. Indeed, CAM-B3LYP, M062x, and B3LYP predict a higher contribution of the tail ligand and a lower one coming from metal and bipyridines.

Concerning the investigation of the triplet excited-state manifold, we decided that for the purpose and relevance of the current work, only the first triplet state, T$_1$, is investigated. From the analysis of the excitation character carried out on various functionals we find that T1 is mostly a tail-to-tail excitation with contributions from the bridging bipyridine. 

In the cases of CAM-B3LYP and M062x we observe that the effect of long-range interactions and higher exact exchange pushed up the bright state, which in these cases is also the S$_1$. In general, it is important to point out that, although more computationally demanding, for a heavy metal like ruthenium, these types of functionals can better handle the metal's electron distribution and strong correlation effects, leading to improved accuracy in describing bond strengths, electronic structure, and reactivity in such large complexes.

For simplicity and because of a satisfactory agreement with the limited experimental data available, we will mainly refer to CAM-B3LYP as our reference exchange correlation functional in this benchmark work. 

\clearpage

\subsubsection*{Geometries}
\lstset{basicstyle=\tiny}
\begin{multicols}{2}
\noindent\textbf{Benzene}
\begin{lstlisting}
    12
charge=0 spin=0 xy-plane
C  0.000000  1.396792  0.000000
C  1.209657  0.698396  0.000000
C  1.209657 -0.698396  0.000000
C  0.000000 -1.396792  0.000000
C -1.209657 -0.698396  0.000000
C -1.209657  0.698396  0.000000
H  0.000000  2.490290  0.000000
H  2.156660  1.245145  0.000000
H  2.156660 -1.245145  0.000000
H  0.000000 -2.490290  0.000000
H -2.156660 -1.245145  0.000000
H -2.156660  1.245145  0.000000
\end{lstlisting}
\textbf{Ferrocene}
\begin{lstlisting}
21
charge=0 spin=0 Cp rings on xy-plane
Fe   0.000000   0.000000   0.000000
C   -0.713500  -0.982049  -1.648000
C    0.713500  -0.982049  -1.648000
C    1.154467   0.375109  -1.648000
C    0.000000   1.213879  -1.648000
C   -1.154467   0.375109  -1.648000
H   -1.347694  -1.854942  -1.638208
H    1.347694  -1.854942  -1.638208
H    2.180615   0.708525  -1.638208
H    0.000000   2.292835  -1.638208
H   -2.180615   0.708525  -1.638208
C   -0.713500  -0.982049   1.648000
C   -1.154467   0.375109   1.648000
C    0.000000   1.213879   1.648000
C    1.154467   0.375109   1.648000
C    0.713500  -0.982049   1.648000
H   -1.347694  -1.854942   1.638208
H   -2.180615   0.708525   1.638208
H    0.000000   2.292835   1.638208
H    2.180615   0.708525   1.638208
H    1.347694  -1.854942   1.638208
\end{lstlisting}
\textbf{Dobna}
\begin{lstlisting}
32
charge=0 spin=0 xz-plane
C -0.070258 -1.218040  3.281650
C  0.000000  0.000000  3.961013
C  0.070258  1.218040  3.281650
C  0.065226  1.196868  1.887643
C  0.000000  0.000000  1.151178
C -0.065226 -1.196868  1.887643
H -0.123911 -2.164118  3.809230
H  0.000000  0.000000  5.047587
H  0.123911  2.164118  3.809230
O -0.133140 -2.404377  1.254718
O  0.133140  2.404377  1.254718
C  0.067297  2.520862 -0.111881
C -0.015330  1.419383 -0.996798
C  0.085074  3.849187 -0.558396
C -0.117094  1.746756 -2.370356
C  0.000000  4.112115 -1.916767
H  0.157462  4.646092  0.174846
C -0.110160  3.054904 -2.831751
H -0.221740  0.945971 -3.093498
H  0.010351  5.140754 -2.267216
H -0.191877  3.260105 -3.895336
C -0.067297 -2.520862 -0.111881
C -0.085074 -3.849187 -0.558396
C  0.015330 -1.419383 -0.996798
C  0.000000 -4.112115 -1.916767
H -0.157462 -4.646092  0.174846
C  0.117094 -1.746756 -2.370356
C  0.110160 -3.054904 -2.831751
H -0.010351 -5.140754 -2.267216
H  0.221740 -0.945971 -3.093498
H  0.191877 -3.260105 -3.895336
B  0.000000  0.000000 -0.370485
\end{lstlisting}
\textbf{Trans-Cl labeled, S0}
\begin{lstlisting}
27
charge=0 spin=0 Ligand placed along z-axis
Ru	 0    1.9888    1.0728
N.0   -1.3313    0.3006    1.0728
C.0   -2.6704    0.3806    1.0728
C.0   -3.4913   -0.7429    1.0728
C.0   -2.8956   -2.0018    1.0728
C.0   -1.5069   -2.0889    1.0728
C.0   -0.7394   -0.9193    1.0728
N.0    1.3313    0.3006    1.0728
C.0    2.6704    0.3806    1.0728
C.0    3.4913   -0.7429    1.0728
C.0    2.8956   -2.0018    1.0728
C.0    1.5069   -2.0889    1.0728
C.0    0.7394   -0.9193    1.0728
H   -3.0914    1.3785    1.0728
H   -4.5692   -0.6221    1.0728
H   -3.4993   -2.9044    1.0728
H   -1.0288   -3.0610    1.0728
H    1.0288   -3.0610    1.0728
H    3.4993   -2.9044    1.0728
H    4.5692   -0.6221    1.0728
H    3.0914    1.3785    1.0728
C.1    1.3191    3.3233    1.0728
C.1   -1.3191    3.3233    1.0728
O    2.1524    4.1192    1.0728
O   -2.1524    4.1192    1.0728
Cl        0    1.8437   -1.3536
Cl        0    1.8437    3.4992

\end{lstlisting}

\textbf{TLD-1433 in H$_2$O labeled, S0}
\begin{lstlisting}
99 
charge=+2, spin=0 
C.0     -5.813102        3.731623       -0.007491
C.0     -4.505283        3.313187        0.006487
C.0     -4.235573        1.938416        0.030146
C.0     -5.329637        1.048678        0.037903
C.0     -6.841183        2.790822        0.004331
C.0     -5.141368       -0.370718        0.062995
C.0     -3.856622       -0.940941        0.082280
C.0     -3.750770       -2.335734        0.105472
H.0     -2.772898       -2.800372        0.120643
C.0     -4.901880       -3.085348        0.107841
C.0     -6.141632       -2.447570        0.085861
H.0     -6.063237        4.784014       -0.026498
H.0     -3.698963        4.036139       -0.000813
H.0     -7.876547        3.103703       -0.001140
H.0     -4.868435       -4.166711        0.125790
H.0     -7.057734       -3.022921        0.083177
N.0     -6.268358       -1.127102        0.061298
N.0     -6.614130        1.483597        0.029337
Ru       -8.017376       -0.030773        0.041759
N.1     -9.284988       -1.645118        0.223198
C.1     -9.856524       -2.310119       -0.786991
C.1     -9.560816       -2.040621        1.484056
C.1    -10.707233       -3.376379       -0.595833
H.1     -9.611597       -1.970394       -1.784184
C.1    -10.410350       -3.107974        1.738255
C.1    -11.006456       -3.802699        0.696016
H.1    -11.133883       -3.871853       -1.459339
H.1    -10.612661       -3.404828        2.758533
N.1     -8.117622       -0.254818        2.089147
C.1     -8.894218       -1.264958        2.534503
C.1     -7.477194        0.495726        2.992565
C.1     -9.036910       -1.524628        3.890353
C.1     -7.580261        0.285392        4.349734
H.1     -6.864910        1.297578        2.602550
C.1     -8.378351       -0.749030        4.833132
H.1     -9.667268       -2.339564        4.219986
H.1     -7.035428        0.932174        5.026533
N.2     -9.664030        1.194200       -0.144275
C.2    -10.394136        1.683758        0.864018
C.2    -10.032122        1.501882       -1.406080
C.2    -11.497522        2.485037        0.670142
H.2    -10.070225        1.421563        1.862141
C.2    -11.135460        2.303346       -1.663093
C.2    -11.897570        2.814522       -0.622721
H.2    -12.042264        2.849839        1.532303
H.2    -11.408134        2.533793       -2.684193
N.2     -8.166370        0.161028       -2.006464
C.2     -9.179858        0.932040       -2.454260
C.2     -7.347370       -0.392464       -2.907855
C.2     -9.379406        1.149149       -3.810360
C.2     -7.496157       -0.212909       -4.265267
H.2     -6.548034       -1.007013       -2.516267
C.2     -8.535583        0.577060       -4.750986
H.2    -10.200183        1.770635       -4.142230
H.2     -6.798044       -0.692621       -4.940268
C.2    -13.094933        3.669710       -0.870648
C.1     -8.513264       -1.006203        6.296744
H.2    -14.003380        3.149533       -0.555098
H.2    -13.194539        3.924072       -1.925259
H.2    -13.036202        4.593088       -0.290423
H.1     -7.539497       -1.239907        6.734417
H.1     -9.190786       -1.835370        6.497204
H.1     -8.889397       -0.117110        6.808556
C.2     -8.725439        0.797025       -6.214541
H.2     -9.615812        1.390849       -6.418038
H.2     -7.860302        1.314500       -6.636988
H.2     -8.814724       -0.157750       -6.737925
C.0     -2.751507       -0.039495        0.076039
C.0     -2.951752        1.333451        0.050143
N.0     -1.423189       -0.331545        0.094398
N.0     -1.699316        1.873003        0.052833
C.0     -0.810549        0.838347        0.080814
C.0      0.615613        1.012340        0.093889
C.1    -11.930470       -4.948222        0.941647
H.1    -11.971848       -5.209888        1.998362
H.1    -11.614470       -5.826634        0.374674
H.1    -12.941351       -4.697991        0.608925
C.0      1.358086        2.168001        0.120119
C.0      2.741257        1.928460        0.132381
C.0      3.060940        0.588514        0.115597
C.0      4.364163       -0.022238        0.123080
C.0      4.694728       -1.349476        0.262599
C.0      6.079761       -1.585540        0.231886
C.0      6.824703       -0.441623        0.067798
C.0      8.258065       -0.296265       -0.004148
C.0      9.011155        0.851981        0.058092
C.0     10.399417        0.612999       -0.045850
C.0     10.694283       -0.711695       -0.185746
S.0      1.632872       -0.375372        0.077817
S.0      5.790692        0.931986       -0.062228
S.0      9.283167       -1.670238       -0.198885
H.0      0.930423        3.161890        0.137951
H.0      3.484014        2.715104        0.163629
H.0      3.958362       -2.131474        0.399072
H.0      6.522257       -2.567260        0.343063
H.0     11.663321       -1.177667       -0.282237
H.0     11.150854        1.390728       -0.013627
H.0      8.578380        1.836160        0.187527
H.0     -1.466427        2.852141        0.034961

\end{lstlisting}
\end{multicols}